\documentclass[12pt,a4paper,notitlepage]{article}

\usepackage{graphicx}
\usepackage{amssymb}
\usepackage{amsmath}

\usepackage{amsthm}
\usepackage{amsfonts}
\usepackage{times}

\usepackage[format=hang]{subcaption}
\usepackage{color}
\usepackage[T1]{fontenc}
\usepackage{hyperref}
\usepackage[square]{natbib}
\usepackage{upgreek}

\usepackage{amsfonts}
\usepackage{pgfplots}
\usepackage{tikz}
\usepackage{tikz-3dplot}
\usepackage{leftidx}
\usepackage{float}
\usepackage{amsmath}
\usepackage{bm}
\usepackage{xcolor}
\AtBeginDocument{}

\date{}

\pgfplotsset{compat=1.14}

\definecolor{rwth1}{RGB}{0,84,159}      % RWTH-Blau
\definecolor{rwth2}{RGB}{142,186,229}   % RWTH-Hellblau
\definecolor{rwth3}{RGB}{0,97,101}      % Petrol 
\definecolor{rwth4}{RGB}{0,152,161}     % Türkis
\definecolor{rwth5}{RGB}{87,171,39}     % Grün
\definecolor{rwth6}{RGB}{189,205,0}     % Maigrün
\definecolor{rwth7}{RGB}{255,237,0}     % Gelb
\definecolor{rwth8}{RGB}{246,168,0}     % Orange
\definecolor{rwth9}{RGB}{227,0,102}     % Magenta
\definecolor{rwth10}{RGB}{204,7,30}     % Rot
\definecolor{rwth11}{RGB}{161,16,53}    % Bordeaux
\definecolor{rwth12}{RGB}{97,33,88}     % Violett
\definecolor{rwth13}{RGB}{122,111,172}  % Lila

\definecolor{sfb1}{RGB}{0,84,165}      % Blau
\definecolor{sfb2}{RGB}{201,0,35}      % Rot
\definecolor{sfb3}{RGB}{231,95,1}      % Orange
\definecolor{sfb4}{RGB}{127,127,127}   % Grau
\definecolor{sfb5}{RGB}{217,217,217}   % Hellgrau

\tikzstyle{dashpattern0} = [dash pattern = ]
\tikzstyle{dashpattern1} = [dash pattern = on 4.25pt off 0.75pt]
\tikzstyle{dashpattern2} = [dash pattern = on 1.5pt off 0.5pt]
\tikzstyle{dashpattern3} = [dash pattern = on 0.75pt off 0.4pt]
\tikzstyle{dashpattern4} = [dash pattern = on 3pt off 1pt on 1pt off 1pt]
\tikzstyle{dashpattern5} = [dash pattern = on 3.75pt off 0.5pt on 0.75pt off 0.5pt on 0.75pt off 0.5pt]
\tikzstyle{dashpattern6} = [dash pattern = on 3.25pt off 0.5pt on 0.75pt off 0.5pt on 0.75pt off 0.5pt on 0.75pt off 0.5pt]
\tikzstyle{dashpattern7} = [dash pattern = on 3.25pt off 0.5pt on 0.75pt off 0.5pt on 0.75pt off 0.5pt on 0.75pt off 0.5pt on 0.75pt off 0.5pt]
\tikzstyle{dashpattern8} = [line cap=round, dash pattern = on 3.25pt off 2.75pt]
\tikzstyle{dashpattern9} = [line cap=round, dash pattern = on 0.01pt off 2pt]
\tikzstyle{dashpattern10}= [line cap=round, dash pattern = on 3.25pt off 2pt on 0.01pt off 2pt]
\tikzstyle{dashpattern11}= [line cap=round, dash pattern = on 3.5pt off 1.75pt on 0.01pt off 1.75pt on 0.01pt off 1.75pt]
\tikzstyle{dashpattern12}= [line cap=round, dash pattern = on 3.5pt off 1.75pt on 0.01pt off 1.75pt on 0.01pt off 1.75pt on 0.01pt off 1.75pt]
\tikzstyle{dashpattern13}= [line cap=round, dash pattern = on 3.5pt off 1.75pt on 0.01pt off 1.75pt on 0.01pt off 1.75pt on 0.01pt off 1.75pt on 0.01pt off 1.75pt]

\newcommand\BbbGammaVar{\reflectbox{\rotatebox[origin=c]{180}{$\mathbb L$}}}

\begin{document}

\author{\large {Christian Gierden$^{a,}$\footnote{Corresponding author: christian.gierden@ifam.rwth-aachen.de} , Johanna Waimann$^{a}$, Bob Svendsen$^{b,c}$, Stefanie Reese$^{a}$}\\[0.5cm]
\hspace*{-0.1cm}
\normalsize{\em $^{a}$ Institute of Applied Mechanics, RWTH Aachen University,
  D-52074 Aachen, Germany}\\
\normalsize{\em $^{b}$ Material Mechanics, RWTH Aachen University, D-52062 Aachen, Germany}\\
\normalsize{\em $^{c}$ Microstructure Physics and Alloy Design, Max-Planck-Institut f\"ur Eisenforschung GmbH}\\
\normalsize{D-40237 D\"usseldorf, Germany}\\
}

\title{\LARGE A geometrically adapted reduced set of frequencies for a FFT-based microstructure simulation}
\maketitle

\small
{\bf Abstract.} {We present a modified model order reduction (MOR) technique for the FFT-based simulation of composite microstructures. It utilizes the earlier introduced MOR technique (\cite{Kochmann19}), which is based on solving the Lippmann-Schwinger equation in Fourier space by a reduced set of frequencies. Crucial for the accuracy of this MOR technique is on the one hand the amount of used frequencies and on the other hand the choice of frequencies used within the simulation. \citet{Kochmann19} defined the reduced set of frequencies by using a fixed sampling pattern, which is most general but leads to poor microstructural results when considering only a few frequencies. Consequently, a reconstruction algorithm based on the $TV_1$-algorithm \citep{Candes06} was used in a post-processing step to generate highly resolved micromechanical fields.\\
The present work deals with a modified sampling pattern generation for this MOR technique. Based on the idea, that the micromechanical material response strongly depends on the phase-wise material behavior, we propose the usage of sampling patterns adapted to the spatial arrangement of the individual phases. This leads to significantly improved microscopic and overall results. Hence, the time-consuming reconstruction in the post-processing step that was necessary in the earlier work is no longer required. To show the adaptability and robustness of this new choice of sampling patterns, several two dimensional examples are investigated. In addition, also the 3D extension of the algorithm is presented.}

\vspace*{0.3cm}

\normalsize

%%%%%%%%%%%%%%%%%%%%%%%%%%%%%%%%%%%%%%%%%%%%%%%%%%%%%%%%%%%%%%%%%%%%%%%%%%%%%%%%%%%%%%%%%%%%%%%%%%%%%%%%%

\section{Introduction}
\label{sec:introduction}
To calculate spatial resolutions for complex microstructural material behaviors within structural finite element (FE) simulations, a two-scale full field simulation is necessary. To perform these highly resolved two-scale simulations, various methodologies have been established (\cite{Geers10}). Examples of these are the FE$^2$ method (e.g. \cite{Smit98, Feyel00}) and the FE-FFT method (e.g. \cite{Spahn14, Kochmann16}). In this context, we focus on the FE-FFT-based simulation approach, but restrict ourselves in this work exclusively to the FFT-based microstructure simulation. Such a microstructure may be given in terms of a representative volume element (RVE) or a unit cell (e.g. \cite{Hill63, Ostoja-Starzewski02}). \\  
The FFT-based modelling of periodic microstructures was introduced by \cite{Mou94, Mou98}. Based on fixed-point iterations, it is used for the simulation of different microstructures, such as composites \citep{Dre00} and polycrystals \citep{Lebensohn01}. In the last two decades improvements of the solution behavior of the FFT-based method were gained in various ways. For example, by the development of more efficient solvers, which are numerically more robust and lead to better convergence behavior. Among these are polarization-based formulations \citep{Ey99, Mon12, Schneider2019}, formulations based on augmented Lagrangians \citep{Mich00, Mich01}, or formulations based on conjugate gradients \citep{Ze10, Bris10, Gele13, Kabel14}. In addition, numerical resolution problems related to the Gibbs phenomenon \citep{Gibbs1898} have been addressed by using e.g. first- \citep{Will14, Willot2015} and higher-order \citep{DKochmann2017} finite difference approximations of the differential operator. In summary, the FFT-based microstructure simulation is an accurate and efficient solution scheme, which is even more efficient than the common FE simulation as shown by \cite{Michel99} or \cite{Pra09}. Nevertheless, the computational effort especially in the context of a two-scale FE-FFT-based simulation is still extremely high. Hence, the development of even more efficient methods is necessary.\\
One possibility is to use hybrid homogenization methods, which combine numerical simulations and theoretical investigations, such as the uniform \citep{Dvorak1992} or non-uniform \citep{Michel2003,Fritzen2010} transformation analysis or the clustering analysis \citep{Liu16,Wulfinghoff2017}. Concerning numerically efficient FE-FFT-based homogenization techniques a straight-forward ansatz is based on using a coarsely discretized microstructure. Consequently, this leads to coarse microstructural results so that a post-processing step is necessary to generate the required highly resolved microstructural fields \citep{Kochmann17, Gierden20}. Other model order reduction techniques are for example based on a proper orthogonal decomposition (POD) \citep{Pinnau2008} using the strain tensor in Fourier space $\hat{\boldsymbol \varepsilon}$ to compute the required projection tensor \citep{Garcia17}, low-rank approximations \citep{Vondrejc19} or on compuations using a reduced set of frequencies \citep{Kochmann19} in Fourier space. \\  
The present paper deals with the model order reduction technique considering a reduced set of frequencies \citep{Kochmann19}. This technique reduces the computational effort of the spectral solver and is suitable for small \citep{Kochmann19} and finite strain kinematics \citep{Gierden19b}. The earlier proposed solution procedure uses a fixed sampling pattern to define the reduced set of frequencies. During the computations, this reduced set of frequencies is used to solve the Lippmann-Schwinger equation in Fourier space. Subsequently, in a post-processing step, a reconstruction and a compatibility step is performed to generate highly resolved microstructural fields. Since the selection of frequencies is crucial for the accuracy of the simulation in our current work, we propose to use a geometrically adapted sampling pattern instead, which incorporates the microstructural phase distribution. This leads to significantly better microstructural results, so that the time consuming reconstruction step using the $TV_1$-algorithm proposed by \cite{Candes06} is not necessary anymore.\\
The paper is structured as follows. The microscopic boundary value problem (BVP) is reviewed in Section \ref{sec:microproblem}. The solution strategy to this microscopic BVP by using a spectral solver and fixed-point iterations is given in Section \ref{sec:spectralsolver}. Section \ref{sec:MORFFT} gives an overview of the recently introduced model order reduction technique based on a reduced set of frequencies with a fixed sampling pattern and the new technique based on using a geometrically adapted sampling pattern. A comparison of the results of both methods is presented in Section \ref{sec:examples}. The paper ends with a conclusion and an outlook in Section~\ref{sec:summary}.\\

\textbf{Notation}. A direct tensor notation is preferred throughout the text. Vectors and second-order tensors are represented by bold letters, e.g. \(\bm{a}\) and \(\bm{A}\), a tensor of fourth order by, e.g., \(\mathbb{A}\). The linear mapping of a second-order tensor \(\bm{B}\) by a fourth-order tensor \(\mathbb{C}\) is denoted by \(\bm{A}=\mathbb{C}:\bm{B}\). The scalar and dyadic products are denoted, e.g. by \(\bm{a}\cdot\bm{b}\) as well as \(\bm{A}:\bm{B}\), and \(\bm{A}\otimes\bm{B}\), respectively. Furthermore, \(||\bullet||\) represents the Frobenius norm and a bar over any quantity $\bar A$ always refers to a macroscopic quantity, while the absence of a bar is related to microscopic quantities. Additional notation is introduced when needed. 

\section{Microscopic boundary value problem (BVP)}
\label{sec:microproblem}
\begin{figure}[H]
	\hspace{-1.5cm}
	\centering
	\includegraphics[width=0.55\textwidth]{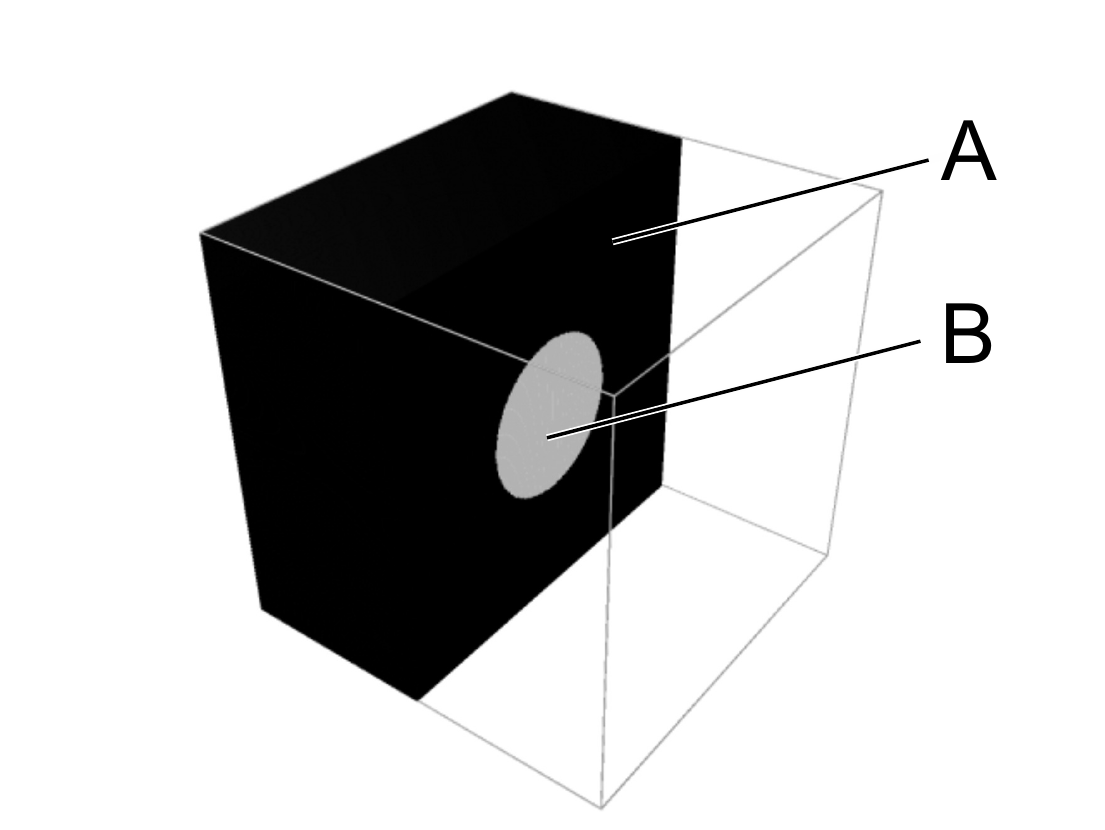}
	\put(-40,151){\colorbox{white}{$\Omega_\mathrm{M}$}}
	\put(-40,109){\colorbox{white}{$\Omega_\mathrm{I}$}}
	\caption{Microstructure with one centered spherical inclusion.}
	\label{fig:Micro}
\end{figure}

Let us consider an inhomogeneous periodic microstructure $\Omega=\Omega_\mathrm{I} \cup \Omega_\mathrm{M}$ with inclusions $\Omega_\mathrm{I}$ embedded in a softer matrix material $\Omega_\mathrm{M}$. One example is a microstructure with one centered spherical inclusion as shown in Figure \ref{fig:Micro}.
Considering small strain kinematics, the total strain $\boldsymbol \varepsilon(\bar{\boldsymbol x}, \boldsymbol x)=\bar{\boldsymbol \varepsilon}(\bar{\boldsymbol x})+\tilde{\boldsymbol \varepsilon}({\boldsymbol x})$ at the macroscopic position $\bar{\boldsymbol x}$ and the microscopic position $\boldsymbol x$ is additively split into the macroscopic part $\bar{\boldsymbol \varepsilon}(\bar{\boldsymbol x})$ and the microscopic fluctuating part $\tilde{\boldsymbol \varepsilon}({\boldsymbol x})$, respectively. Subjecting the given microstructure to a macroscopic strain $\bar{\boldsymbol \varepsilon}(\bar{\boldsymbol x})$, the total stress $\boldsymbol \sigma(\bar{\boldsymbol x}, \boldsymbol x)$ is computed by solving the microscopic boundary value problem 
\begin{align}
\begin{array}{rcl}
\text{div}\,\boldsymbol \sigma(\bar{\boldsymbol x},\boldsymbol x) &=& \boldsymbol 0 \qquad\qquad\qquad\qquad\qquad \forall \boldsymbol x \in \Omega\\
\boldsymbol \sigma(\bar{\boldsymbol x}, \boldsymbol x)&= & \boldsymbol\sigma(\bar{\boldsymbol x}, \boldsymbol x,\boldsymbol \varepsilon(\bar{\boldsymbol x}, \boldsymbol x), \boldsymbol \alpha (\boldsymbol x))\\
\boldsymbol \varepsilon(\bar{\boldsymbol x}, \boldsymbol x)&=&\bar{\boldsymbol \varepsilon}(\bar{\boldsymbol x})+\nabla^S(\tilde{\boldsymbol u}(\boldsymbol x)) 
\end{array}
\label{eq:inhomoBVP}
\end{align}
for each macroscopic position $\bar{\boldsymbol x}$, while body forces are considered only on the macroscopic level and are thus neglected on the micro level. Within Equation \eqref{eq:inhomoBVP}, $\boldsymbol \alpha(\boldsymbol x)$ describes a set of internal variables, $\tilde{\boldsymbol u}(\boldsymbol x)$ is the microscopic fluctuating displacement field and $\nabla^S$ represents the symmetric gradient operator. In regard of comprehensibility, the dependence of all variable besides the macroscopic strains on the macroscopic position $\bar{\boldsymbol x}$ is not shown. The corresponding macroscopic stress $\bar{\boldsymbol \sigma}(\bar{\boldsymbol x})$ and strain $\bar{\boldsymbol \varepsilon}(\bar{\boldsymbol x})$ tensors are defined by the volume average of their local quantities:
\begin{align}
\bar{\boldsymbol \sigma}(\bar{\boldsymbol x}):=\frac{1}{V}\int_\Omega \boldsymbol \sigma (\boldsymbol x) \, \mathrm d\Omega \quad \text{and} \quad \bar{\boldsymbol \varepsilon}(\bar{\boldsymbol x}):=\frac{1}{V}\int_\Omega \boldsymbol \varepsilon (\boldsymbol x) \, \mathrm d\Omega \,.
\end{align}
We restrict ourself to microstructures consisting of two phases with linear elastic or linear elasto-plastic material behavior. Doing that, the total strain can be additively split into an elastic part ${\boldsymbol \varepsilon_\mathrm{e}}({\boldsymbol x})$ and a plastic part ${\boldsymbol \varepsilon_\mathrm{p}}({\boldsymbol x})$: ${\boldsymbol \varepsilon}({\boldsymbol x})={\boldsymbol \varepsilon_\mathrm{e}}({\boldsymbol x})+{\boldsymbol \varepsilon_\mathrm{p}}({\boldsymbol x})$. The linear-elastic stress-strain relation reads $\boldsymbol \sigma(\boldsymbol x)=\mathbb C(\boldsymbol x):\boldsymbol \varepsilon_\mathrm{e} (\boldsymbol x)$. The yield condition for the elasto-plastic behavior is defined as
\begin{align}
\Phi(\boldsymbol \sigma(\boldsymbol x), \varepsilon_\mathrm{p}^{\text{acc}}(\boldsymbol x), \boldsymbol x)=\sigma^{\text{eq}}(\boldsymbol x)-[\sigma_\mathrm{y}^0(\boldsymbol x)+\text H(\boldsymbol x)\,\varepsilon_\mathrm{p}^{\text{acc}}(\boldsymbol x)]
\end{align}
which is the classical von Mises yield condition with an isotropic linear hardening. Here, $\sigma_{\mathrm{y}}^0(\boldsymbol x)$ is the initial yield stress, $\sigma^{\text{eq}}(\boldsymbol x)$ is the von Mises equivalent stress, $H(\boldsymbol x)$ is the hardening modulus, and $\varepsilon_\mathrm{p}^{\text{acc}}(\boldsymbol x)$ is the accumulated plastic strain. In terms of an associative flow rule, the evolution of the plastic strain $\dot{\boldsymbol \varepsilon}_\mathrm{p}$ is given by
\begin{align}
\dot{\boldsymbol \varepsilon}_\mathrm{p}= \dot\gamma \frac{\mathrm{\partial} \Phi}{\mathrm{\partial} \boldsymbol \sigma}
\end{align}
with the plastic multiplier $\dot \gamma$. Finally, the Karush-Kuhn-Tucker conditions $\Phi \le 0$, $\dot \gamma \ge 0$ and $\Phi \, \dot \gamma = 0$ need to be fulfilled.

\section{FFT-based microstructure simulation using the basic fixed-point scheme}
\label{sec:spectralsolver}
To solve the inhomogeneous microscopic boundary value problem introduced in Equation \ref{eq:inhomoBVP}, \citet{Hashin62} proposed to transfer it into an equivalent homogeneous representation 
\begin{align}
\begin{array}{rrl}
\text{div}\,\mathbb C^0 : \boldsymbol  \varepsilon(\boldsymbol x) + \text{div}\,\boldsymbol \tau(\boldsymbol x)&=& \boldsymbol 0 \qquad\qquad\qquad\qquad\qquad\qquad\qquad \forall \boldsymbol x \in \Omega\\
\boldsymbol \tau(\boldsymbol x)&:= & \boldsymbol\sigma(\boldsymbol x,\boldsymbol \varepsilon(\boldsymbol x), \boldsymbol \alpha (\boldsymbol x))- \mathbb C^0 : \boldsymbol \varepsilon (\boldsymbol x)\\
\boldsymbol \varepsilon(\boldsymbol x)&=&\bar{\boldsymbol \varepsilon}(\bar{\boldsymbol x})+\nabla^S(\tilde{\boldsymbol u}(\boldsymbol x)) \,,
\end{array}
\label{eq:homoBVP}
\end{align}
by defining the polarization stress $\boldsymbol \tau (\boldsymbol x)$. It describes the fluctuation of the microstructural stress around the stress of a homogeneous reference material with stiffness $\mathbb C^\mathrm{0}$. Introducing Green's function $\boldsymbol G^0(\boldsymbol x, \boldsymbol x')$ and Green's operator $\BbbGammaVar^{(0)}(\boldsymbol x, \boldsymbol x')$, respectively, the integral equation
\begin{align}
\boldsymbol \varepsilon (\boldsymbol x)= \bar{\boldsymbol \varepsilon}(\bar{\boldsymbol x}) - \int_\Omega \BbbGammaVar^{(0)}(\boldsymbol x, \boldsymbol x^\prime) : \boldsymbol \tau(\boldsymbol x^\prime)\,\text d\boldsymbol x^\prime 
\label{eq:LSE}
\end{align}
enables the solution of Equation \ref{eq:homoBVP}. Equation \ref{eq:LSE} is also known as Lippmann-Schwinger equation [\citet{Kroener1959, Willis81}]. The present convolution intregal is solved by transferring the Lippmann-Schwinger equation into Fourier space yielding
\begin{align}
\hat {\boldsymbol \varepsilon}(\boldsymbol \xi) =  
\left\{ 
\begin{array}{lcl} 
-\hat \BbbGammaVar^{(0)}(\boldsymbol \xi) \,\hat {\boldsymbol \tau} (\boldsymbol \xi) & \forall & \boldsymbol \xi \ne \boldsymbol 0\\
\bar{\boldsymbol\varepsilon} & \forall & \boldsymbol \xi = \boldsymbol 0
\end{array} 
\right.
\label{Eq:LippmannSchwinger}
\end{align}
with the wave vector $\boldsymbol \xi$. In Fourier space, also Green's function and Green's operator 
\begin{align}
\hat G ^0_{ik}(\boldsymbol \xi) &=(\mathbb C^0_{ijkl} \, \xi_j \, \xi_l)^{-1} \\
\text{and} \quad \hat \Gamma_{ijkl}^0(\boldsymbol \xi) &= \frac{1}{4}\left(\hat { G}_{jk,li}^0(\boldsymbol \xi) +\hat { G}_{ik,lj}^0(\boldsymbol \xi) +\hat { G}_{jl,ki}^0(\boldsymbol \xi) +\hat { G}_{il,kj}^0 (\boldsymbol \xi) \right)
\label{eq:Gamma}
\end{align}
are explicitly known. An iterative solution scheme based on Equation \ref{Eq:LippmannSchwinger} was first introduced by \cite{Mou94, Mou98}. The mechanical equilibrium is considered to be achieved when the convergence criterion
\begin{align}
\dfrac{||\bm{\varepsilon}^{(i)}(\bm{x})-\bm{\varepsilon}^{(i-1)}(\bm{x})||}{|\overline{\bm{\varepsilon}}|}<\mathrm{tol}_{\varepsilon}
\end{align}
is fullfilled within iteration step $i$. Considering a fixed-point iteration scheme, the best convergence behavior is achieved by defining the homogeneous reference material behavior based on the arithmetic average of the spatially varying Lam\'{e} constants $\lambda(\boldsymbol x)$ and $\mu(\boldsymbol x)$ \citep{Mou98,Kabel14}. This so-called basic fixed-point scheme is used within all following simulations. In addition we use a first-order finite difference approximation of the differential operator in Equation \ref{eq:Gamma} to avoid numerical resultion problems related to the Gibbs phenomenon, as proposed by \cite{Willot2015}.

\section{Model order reduction technique based on a reduced set of frequencies}
\label{sec:MORFFT}
\begin{figure}[H]
	\centering
	\includegraphics[width=0.9\textwidth]{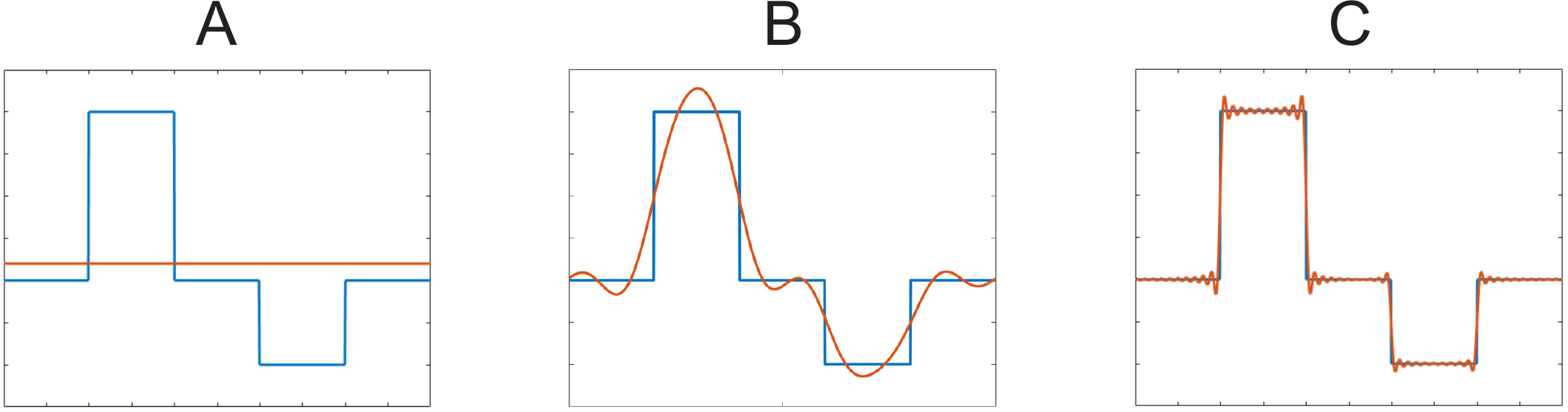}
	\put(-387,100){\colorbox{white}{$\leftidx^R\xi=\{\xi=0\}$}}
	\put(-250,100){\colorbox{white}{$\leftidx^R\boldsymbol \xi=\{\xi_1, ..., \xi_{10}\}$}}
	\put(-105,100){\colorbox{white}{$\leftidx^R\boldsymbol \xi=\{\xi_1, ..., \xi_{100}\}$}}
	\caption{Approximating a step function with a Fourier series consisting of one, ten or 100 frequencies.}
	\label{fig:Frequenzen}
\end{figure}

Recently, \cite{Kochmann19} proposed a model order reduction technique based on a reduced set of wave vectors $\leftidx^R\boldsymbol \xi$ to decrease the computational effort of the spectral solver. The general idea is that any function, such as the step function in Figure \ref{fig:Frequenzen}, may be approximated by a Fourier series. Using a reduced set of frequencies with only one or ten frequencies leads to a very coarse approximation (see Figure \ref{fig:Frequenzen}). Nevertheless, using 100 frequencies already yields an accurate approximation of the step function. More frequencies would lead to even better solutions, but the computational effort will also rise with more frequencies. Due to that, a reduced set of frequencies must be defined in a way, that leads to accurate results but also low computational costs. 
Using this idea in terms of the FFT-based method yields a reduced solution of the Lippmann-Schwinger equation in Fourier space. The resulting reduced fixed-point scheme is given in the following:\\\\
\noindent\textbf{while} \quad $\dfrac{||{\boldsymbol \varepsilon}^{(i+1)}(\boldsymbol x)-{\boldsymbol \varepsilon}^{(i)}(\boldsymbol x)||_{L_2}}{||\bar{\boldsymbol \varepsilon}(\bar{\boldsymbol x})||_{L_2}} \le \text{tol}_{\boldsymbol \varepsilon} $ \quad\textbf{do}
\begin{itemize}
	\item[a)] ${\boldsymbol\tau}^{(i)}(\boldsymbol x)={\boldsymbol \sigma}({\boldsymbol \varepsilon}^{(i)}(\boldsymbol x))-\mathbb C^{0} : \,{\boldsymbol \varepsilon}^{(i)}(\boldsymbol x) \qquad \forall \boldsymbol x \in \Omega$
	\vspace{0.3cm}
	\item[b)] ${\hat{\boldsymbol \tau}}^{(i)}({\boldsymbol \xi}) = \text{{FFT}}\left\{{\boldsymbol \tau}^{(i)}(\boldsymbol x)\right\}$
	\item[c)] $\leftidx{^R}{\hat{\boldsymbol \varepsilon}}^{(i+1)}(\leftidx{^R}{\boldsymbol \xi})=\left\{ \begin{array}{lcl} -\leftidx{^R}{\hat{\BbbGammaVar}^{(0)}}(\leftidx{^R}{\boldsymbol \xi}):\leftidx{^R}{\hat{\boldsymbol \tau}}^{(i)}(\leftidx{^R}{\boldsymbol \xi})&\text{for}& \leftidx{^R}{\boldsymbol \xi} \ne \boldsymbol 0 \\ \hspace{0.5cm} \bar{\boldsymbol \varepsilon}(\bar{\boldsymbol x}) &\text{for}& \leftidx{^R}{\boldsymbol \xi} = \boldsymbol 0\end{array} \right.$
	\item[d)] $\leftidx{^R}{\boldsymbol \varepsilon}^{(i+1)}(\boldsymbol x)={\text{iFFT}}\color{black}\left\{\leftidx{^R}{\hat{\boldsymbol \varepsilon}}^{(i+1)}(\leftidx{^R}{\boldsymbol \xi}) \right\}$
	\vspace{0.2cm}
\end{itemize}
\textbf{end do}\\\\
The speed-up in this algorithm is  gained from step c) by solving the convolution integral in Fourier space using the reduced set of wave vectors. It is also possible to use the reduced set of frequencies and a discrete Fourier transformation (DFT) for the Fourier transformation in step b) and and the inverse Fourier transformation in step d), but since the FFT with the full set of frequencies is even faster than the DFT with the reduced set of frequencies, the FFT is preferably used.\\
Crucial for the performance of this algorithm is the definition of the reduced set of frequencies. In the following, a short review of the recently proposed fixed sampling pattern for the choice of a reduced set of frequencies is given. Subsequently, a new method for the generation of sampling patterns based on the microstructural geometry is presented. For simplicity, both methods are presented for the 2D case.

\subsection{Reduced set of frequencies based on a fixed sampling pattern}
\begin{figure}[H]
	\centering
	\includegraphics[width=\textwidth]{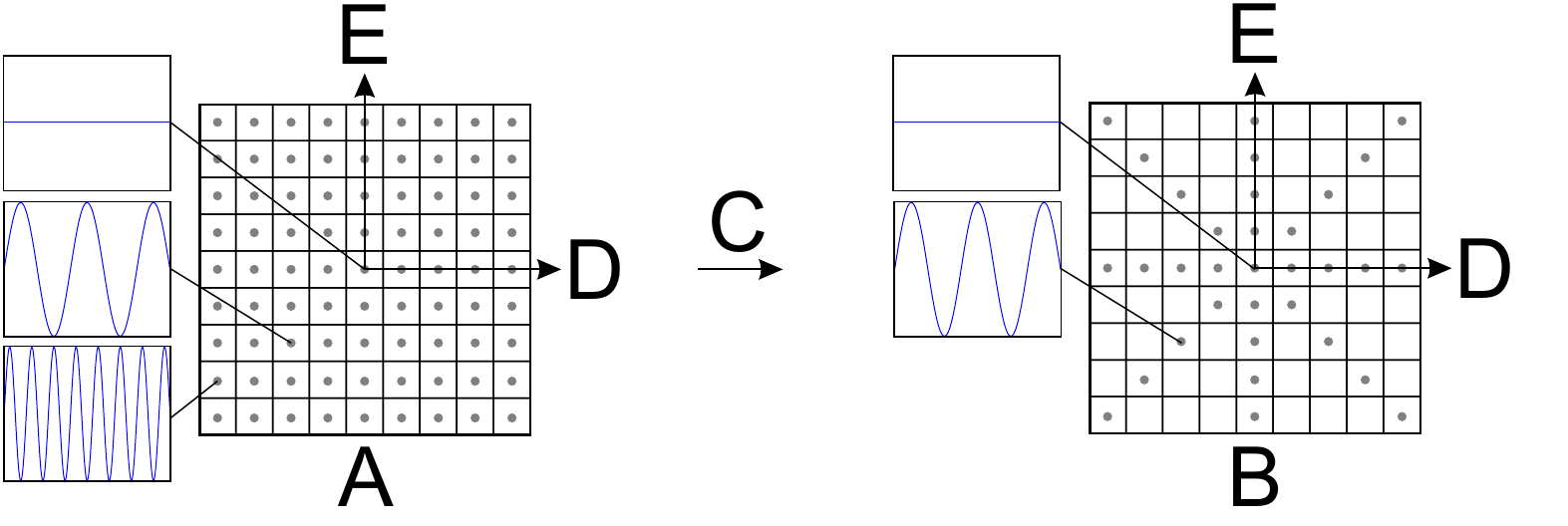}
	\put(-385,11){\colorbox{white}{unreduced set$\phantom{\leftidx^R\xi=\{\xi=0\}}$}}
	\put(-119,11){\colorbox{white}{reduced set$\phantom{\leftidx^R\xi=\{\xi=0\}}$}}
	\put(-298,85){\colorbox{white}{$\phantom{\leftidx^R\xi}$sampling pattern}}
	\put(-292,71){\colorbox{white}{$\xi_1$$\phantom{\leftidx^R\xi}$}}
	\put(-32,71){\colorbox{white}{$\xi_1$$\phantom{\leftidx^R\xi}$}}
	\put(-358,139.2){\colorbox{white}{$\xi_2$$\phantom{\leftidx^R\xi}$}}
	\put(-98,139.2){\colorbox{white}{$\xi_2$$\phantom{\leftidx^R\xi}$}}
	\caption{Unreduced set of frequencies (\textit{left}) and a reduced set of frequencies based on a radial sampling pattern (\textit{right}).}
	\label{fig:FixedSampling}
\end{figure}
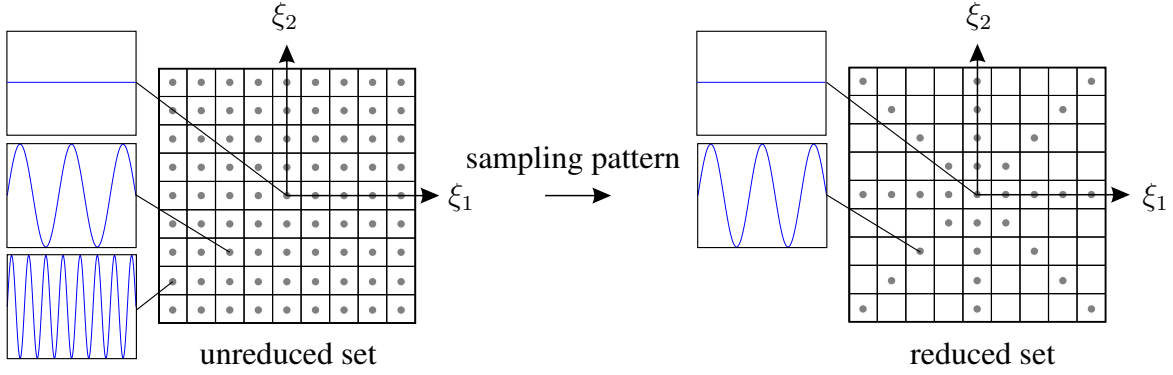

The proposed algorithm by \cite{Kochmann19} is based on the identification of a fixed sampling pattern, which is used to define the reduced set of wave vectors for the solution of the Lippmann-Schwinger equation. This sampling is always the same and therefore does not take the material behavior or the microstructural geometry into account. After performing the online computations with this reduced set of frequencies, a reconstruction based on the $TV_1$-algorithm proposed by \cite{Candes06} is performed. Subsequently, a compatibility step is necessary to generate highly resolved micromechanical fields. To identify an appropriate set of wave vectors, a circular, a squared and a radial sampling pattern are investigated. The most general reduced set of frequencies, which is also most suitable for the reconstruction algorithm, is the radial sampling pattern, shown in Figure \ref{fig:FixedSampling}.
Figure \ref{fig:FixedSampling} shows the full set of frequencies (\textit{left}) as well as a reduced set of frequencies based on a radial sampling pattern (\textit{right}) while the lowest frequencies are in the middle of the shown grid, beginning with the zeroth frequency, which is the mean of the approximated function. \\
If only a few percent of frequencies are considered, the usage of such a reduced set of frequencies leads to poor microstructural and overall results. Due to that, a post-processing step is used after the reduced simulation. This generates highly resolved data by using the $TV_1$-algorithm proposed by \cite{Candes06}. In addition, thereafter, a compatibility step is needed to guarantee a compatible strain field.

\subsection{A novel approach for a reduced set of frequencies based on a geometrically adapted sampling pattern}
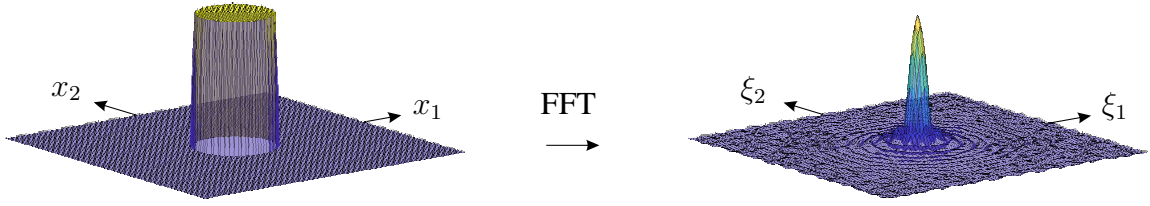
\begin{figure}[H]
	\centering
	\includegraphics[width=0.95\textwidth]{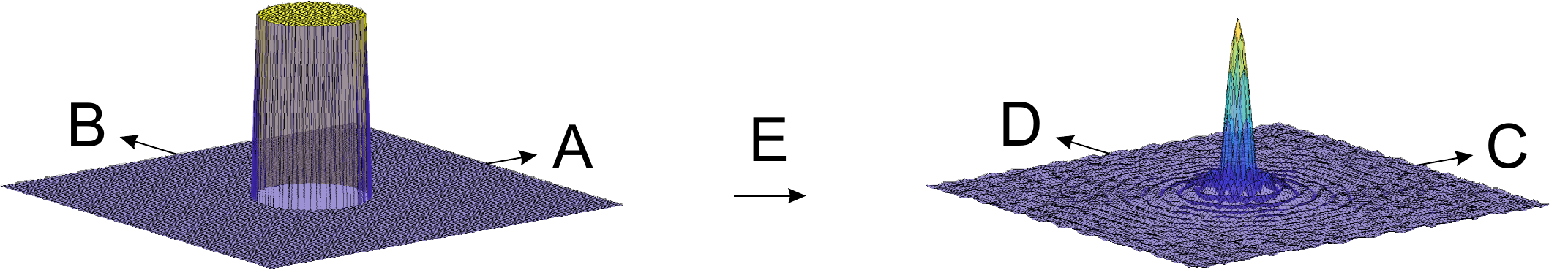}
	\put(-431,39){\colorbox{white}{$\phantom{\leftidx^R\xi}$$x_2$}}
	\put(-283,32){\colorbox{white}{$x_1$$\phantom{\leftidx^R\xi}$}}
	\put(-173,38){\colorbox{white}{$\phantom{\leftidx^R\xi}$$\xi_2$}}
	\put(-25,32){\colorbox{white}{$\xi_1$$\phantom{\leftidx^R\xi}$}}
	\put(-235,32){\colorbox{white}{FFT$\phantom{\leftidx^R\xi}$}}
	\caption{Geometry in real space $g(\boldsymbol x)$ (\textit{left}) and geometry transferred into Fourier space $\hat g(\boldsymbol \xi)$ (\textit{right}).}
	\label{fig:GeoRealFourier}
\end{figure}

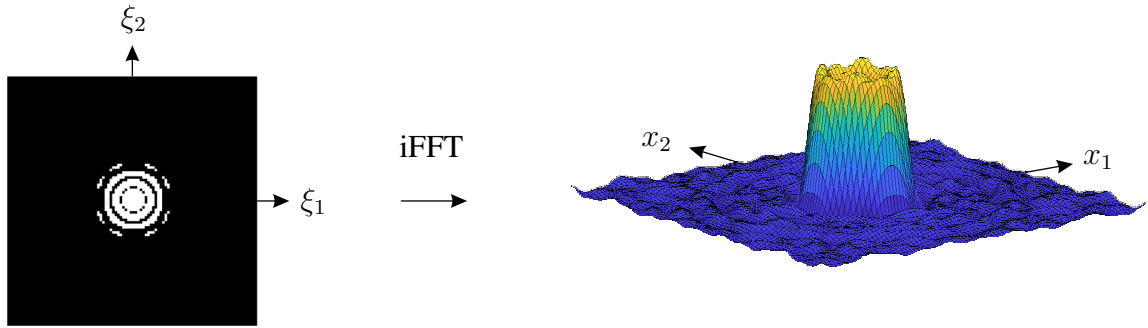
\begin{figure}[H]
	\centering
	\includegraphics[width=0.95\textwidth]{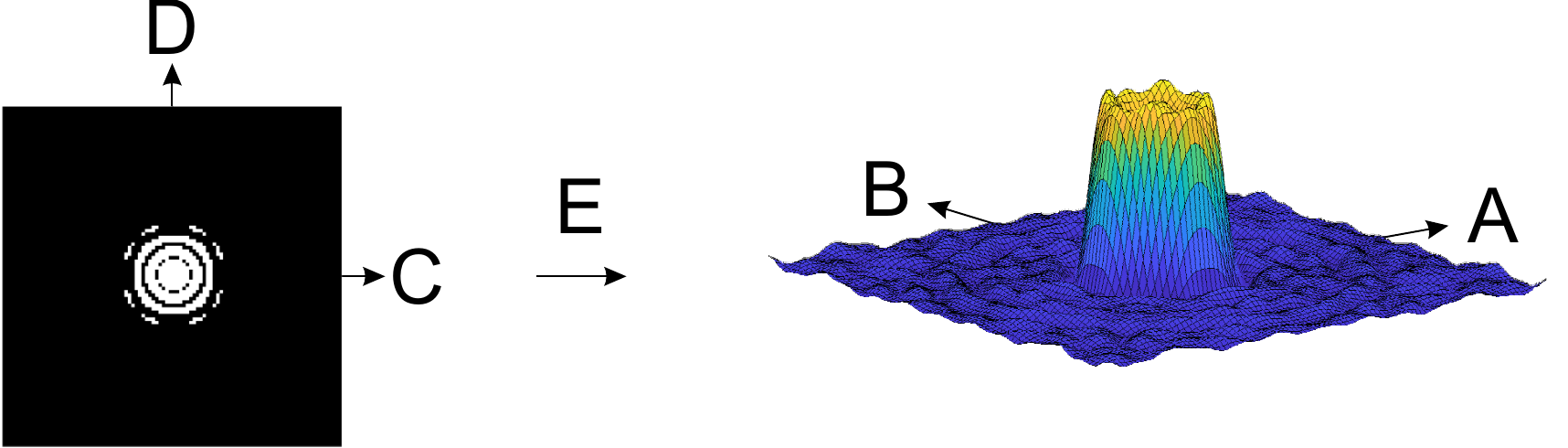}
	\put(-405,113){\colorbox{white}{$\phantom{\leftidx^R\xi}$$\xi_2$}}
	\put(-325,44){\colorbox{white}{$\xi_1$$\phantom{\leftidx^R\xi}$}}
	\put(-210,68){\colorbox{white}{$\phantom{\leftidx^R\xi}$$x_2$}}
	\put(-32,61){\colorbox{white}{$x_1$$\phantom{\leftidx^R\xi}$}}
	\put(-288,64){\colorbox{white}{iFFT$\phantom{\leftidx^R\xi}$}}
	\caption{Geometrically adapted sampling pattern with $\mathcal R=2\,\%$ of frequencies (\textit{left}) and geometrical step function approximated by this reduced set of frequencies (\textit{right}).}
	\label{fig:SamplingReduced}
\end{figure}

The mechanical behavior strongly depends on the geometrical representation of e.g. inclusions within the microstructure. As an example, smaller strains occur in stiffer inclusions and corresponding higher strains in softer matrix material. The idea of an adapted sampling pattern is to use the microstructural geometry to define the reduced set of frequencies. Therefore, the geometry of a two phase material is presented by a step function 
\begin{align}
g(\boldsymbol x) =\begin{cases}
0 & \text{for } \boldsymbol x \in \Omega_\mathrm M\\
1 & \text{for } \boldsymbol x \in \Omega_\mathrm I\,,
\end{cases}
\end{align}
in real space as shown in Figure \ref{fig:GeoRealFourier} (\textit{left}) (with $100 \times 100$ grid points). Transferring this representation of the microstructural geometry into Fourier space results in the plot given in Figure \ref{fig:GeoRealFourier} (\textit{right}), in which the frequencies with the corresponding amplitudes are plotted. Again, the lowest frequencies are centered.
The highest amplitudes, given in this Fourier representation, correspond to the most needed frequencies for the approximation of the geometry. So, a reduced set of frequencies should be defined by taking into account a given percentage of frequencies with the highest amplitudes to obtain the best approximation. Using for example a reduced set of frequencies with $\mathcal R=2\,\%$ of the highest frequencies leads to the sampling pattern given in Figure \ref{fig:SamplingReduced} (\textit{left}), while $\mathcal R$ describes the percentage of used frequencies. The consideration of these frequencies for the approximation of the geometry in real space results in the approximated step function in Figure \ref{fig:SamplingReduced} (\textit{right}).
The idea of a geometrically adapted sampling pattern is to use the same sampling pattern as shown in Figure \ref{fig:SamplingReduced} (\textit{left}) also for the approximation of the microstructural strains and therefore also for solving the Lippmann-Schwinger equation in Fourier space. Doing that, significantly better microscopic and overall results are achieved compared to the fixed radial sampling pattern (see Figure \ref{fig:FixedSampling} (\textit{right})), as shown in Section \ref{sec:examples}. Due to that, the time consuming reconstruction algorithm which was necessary for the fixed sampling pattern is no longer needed. 

\section{Numerical examples and comparison of the results with fixed and adapted sampling patterns}
\label{sec:examples}
To test the adaptivity and accuracy of a geometrically adapted sampling pattern, first, several 2D two phase microstructures with elastic material behavior and one centered inclusion are considered and compared to the results generated by the fixed sampling pattern in Section \ref{chap:OneInc}. This is followed by considering a microstructure with several inclusions and an elastic or elasto-plastic material behavior in Section \ref{chap:several} and the straight forward extension to the 3D case in Section \ref{chap:3D}.\\
All microstructures are assumed to be squared or cubic and discretized by $n=256\times256$ and $n=256\times256\times256$ equidistant grid points, respectively. The mechanical equilibrium is considered to be achieved for the tolerance tol$_\varepsilon<10^{-8}$. To compare the results to the reference solution with the full set of frequencies, a macroscopic error $\bar{\mathcal E}$ and a microscopic error $\mathcal E$ is defined by:
\begin{align*}
\bar{\mathcal E}=\frac{||\bar{\boldsymbol \sigma}-\bar{\boldsymbol \sigma}_{\text{ref}}||}{||\bar{\boldsymbol \sigma}_{\text{ref}}||} \quad \text{and} \quad {\mathcal E}=\frac{1}{n}\sum\limits_n  \frac{||{\boldsymbol \sigma(n)}-{\boldsymbol \sigma}_{\text{ref}}(n)||}{||{\boldsymbol \sigma}_{\text{ref}}(n)||} \,,
\end{align*}
where $n$ is the total number of grid points. All computations are performed using MATLAB with the build-in FFTW-library and pre-compiled material routines on a Dell notebook with a $2.80$ GHz Intel $i7$ quad-core processor and $32$ GB of RAM.

\subsection{Elastic 2D two phase materials with one inclusion}
\label{chap:OneInc}
\begin{figure}[H]
	\centering
	\includegraphics[width=\textwidth]{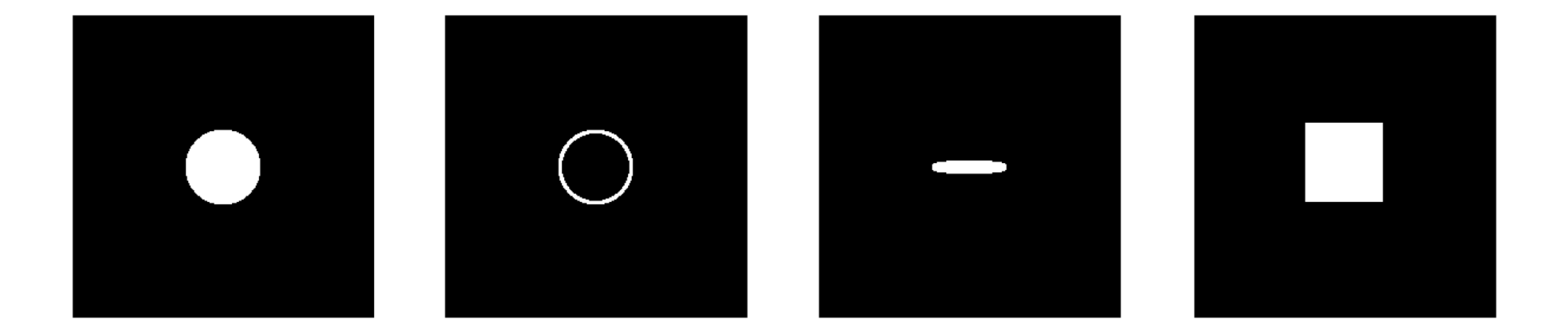}
	\caption{Microstructures with various kinds of central inclusions. The matrix material is colored black and the inclusion is colored white.}
	\label{fig:OneInc}
\end{figure}

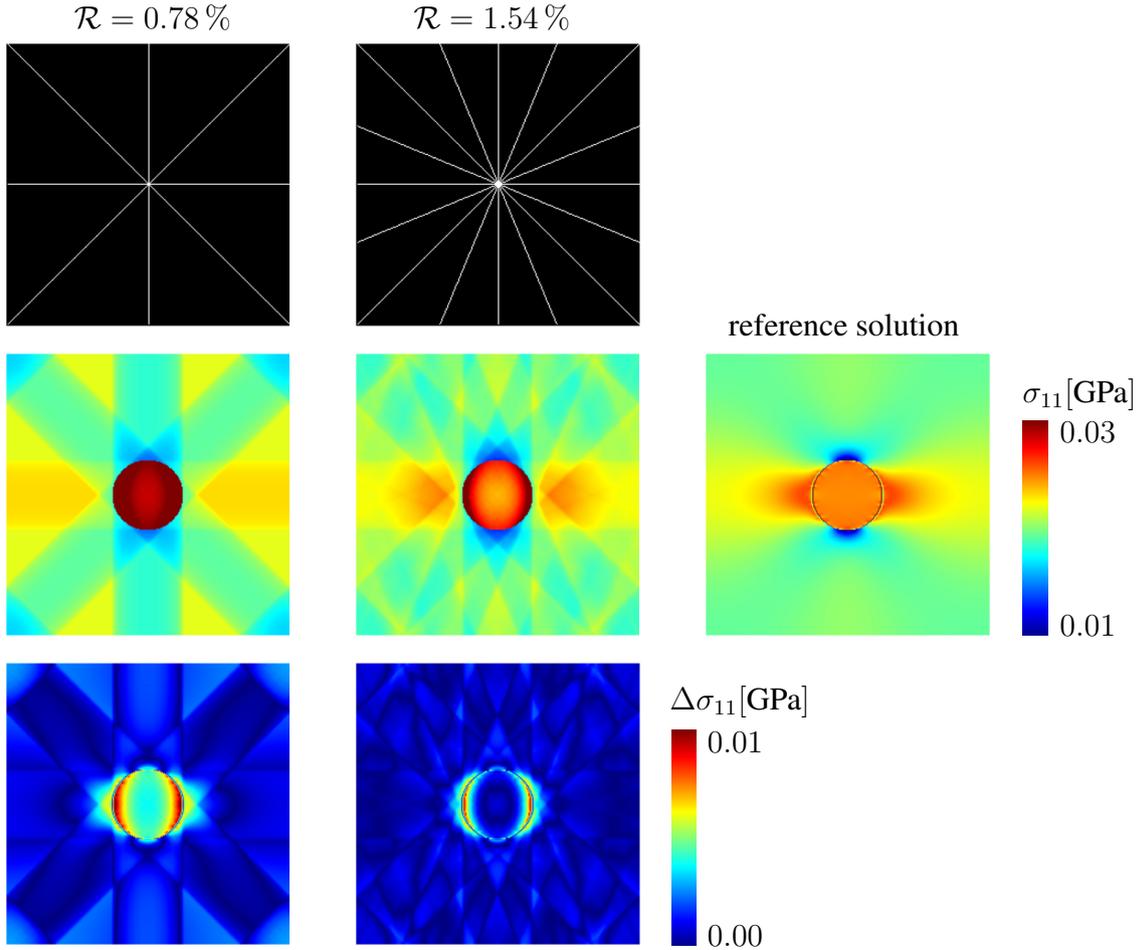
\begin{figure}[H]
	\hspace{-1.5cm}
	\centering
	\includegraphics[width=0.95\textwidth]{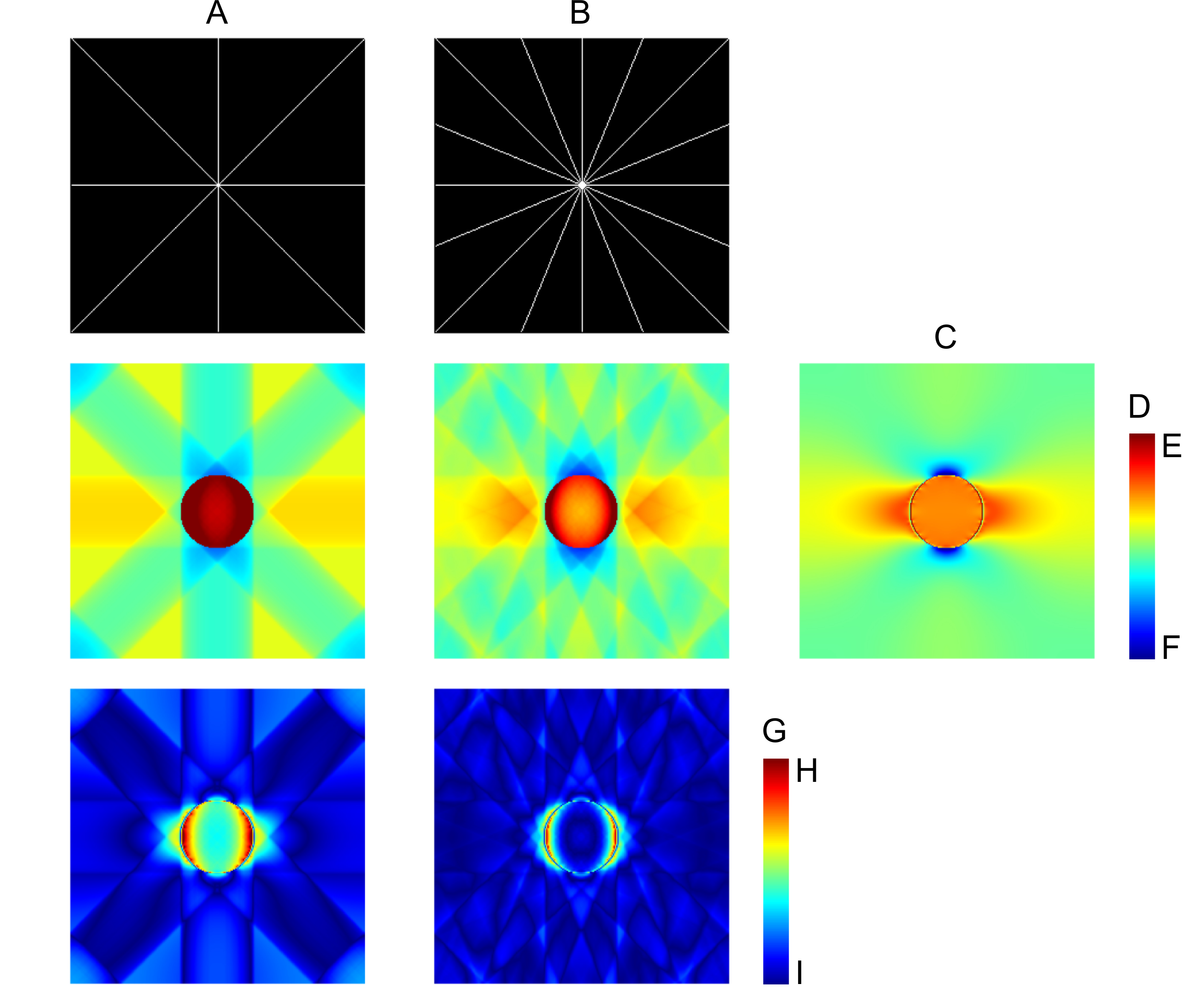}
	\put(-385,351){\colorbox{white}{$\mathcal R = 0.78\,\%$}}
	\put(-258,351){\colorbox{white}{$\mathcal R = 1.54\,\%$}}
	\put(-140,235){\colorbox{white}{reference solution}}
	\put(-30,210){\colorbox{white}{$\sigma_{11}[\text{GPa}]$}}
	\put(-16,195){\colorbox{white}{$0.03$}}
	\put(-16,122){\colorbox{white}{$0.01$}}
	\put(-162,94){\colorbox{white}{$\Delta\sigma_{11}[\text{GPa}]$}}
	\put(-148,78){\colorbox{white}{$0.01$}}
	\put(-148,5){\colorbox{white}{$0.00$}}
	\caption{Microstructural fields for the 2D elastic microstructure with one circular inclusion. \textit{Top row}: Fixed sampling pattern with two different numbers of wave vectors. \textit{Middle row}: Corresponding microstructural stress field $\sigma_{11}$ and reference stress field computed with the full set of frequencies. \textit{Bottom row}: Absolute difference in the microstructural stress field $\Delta\sigma_{11}$.}
	\label{fig:OneCircFixed}
\end{figure}

\begin{figure}[H]
	\hspace{-1.5cm}
	\centering
	\includegraphics[width=0.95\textwidth]{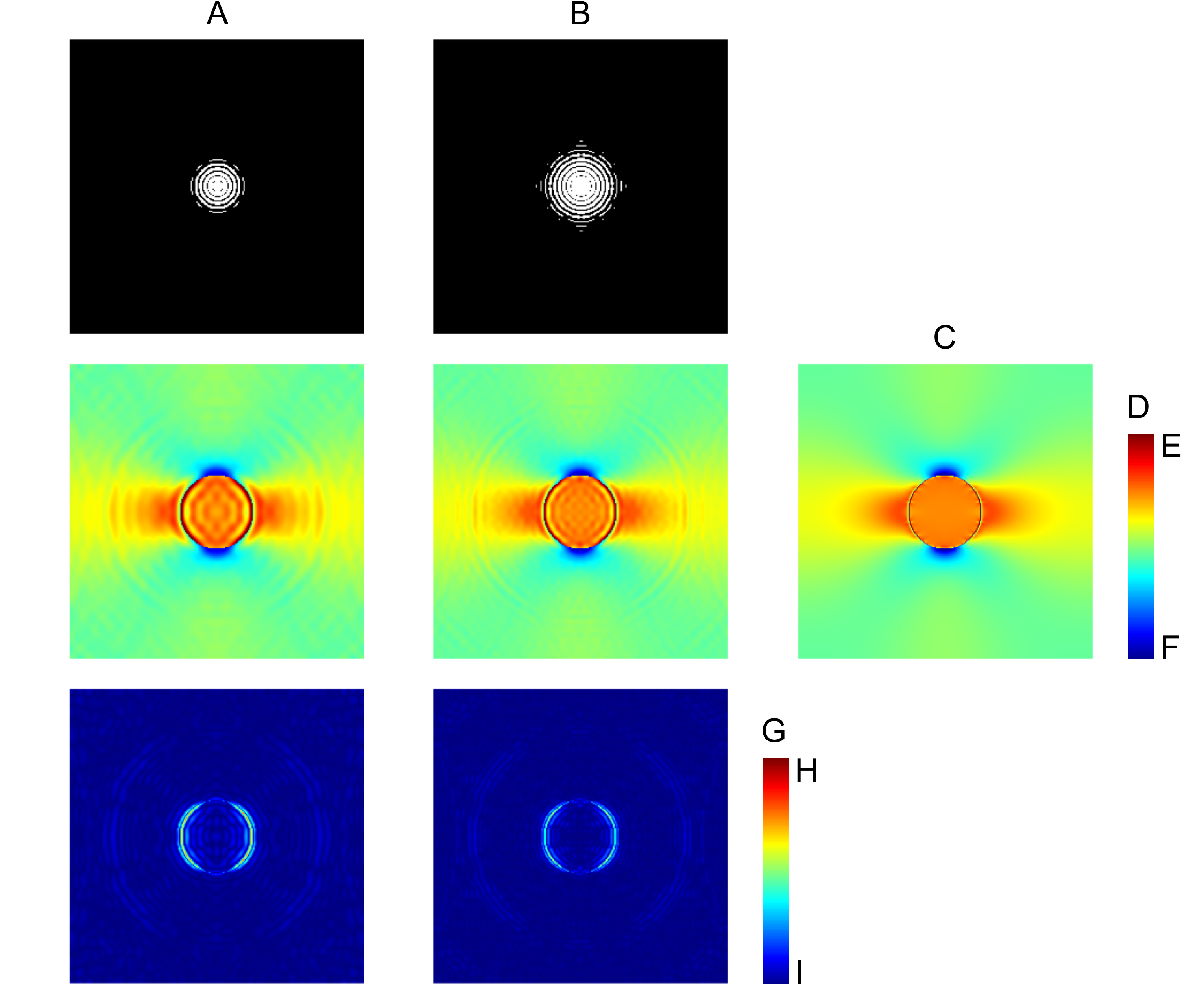}
	\put(-385,351){\colorbox{white}{$\mathcal R = 0.78\,\%$}}
	\put(-258,351){\colorbox{white}{$\mathcal R = 1.54\,\%$}}
	\put(-140,235){\colorbox{white}{reference solution}}
	\put(-30,210){\colorbox{white}{$\sigma_{11}[\text{GPa}]$}}
	\put(-16,195){\colorbox{white}{$0.03$}}
	\put(-16,122){\colorbox{white}{$0.01$}}
	\put(-162,94){\colorbox{white}{$\Delta\sigma_{11}[\text{GPa}]$}}
	\put(-148,78){\colorbox{white}{$0.01$}}
	\put(-148,5){\colorbox{white}{$0.00$}}
	\caption{Microstructural fields for the 2D elastic microstructure with one circular inclusion. \textit{Top row}: Adapted sampling pattern with two different numbers of wave vectors. \textit{Middle row}: Corresponding microstructural stress field $\sigma_{11}$ and reference stress field computed with the full set of frequencies. \textit{Bottom row}: Absolute difference in the microstructural stress field $\Delta\sigma_{11}$.}
	\label{fig:OneCircAdapted}
\end{figure}
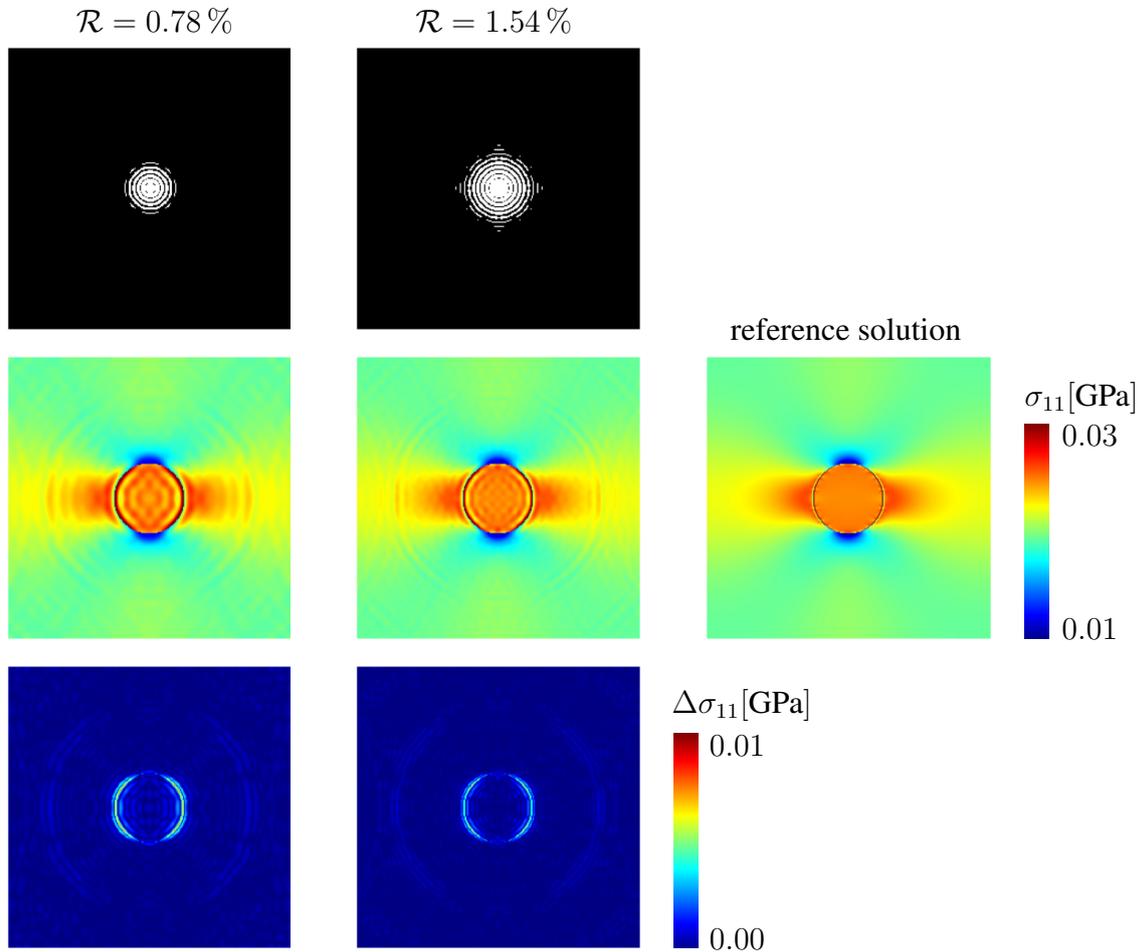

\begin{figure}[H]
	\includegraphics[width=\textwidth]{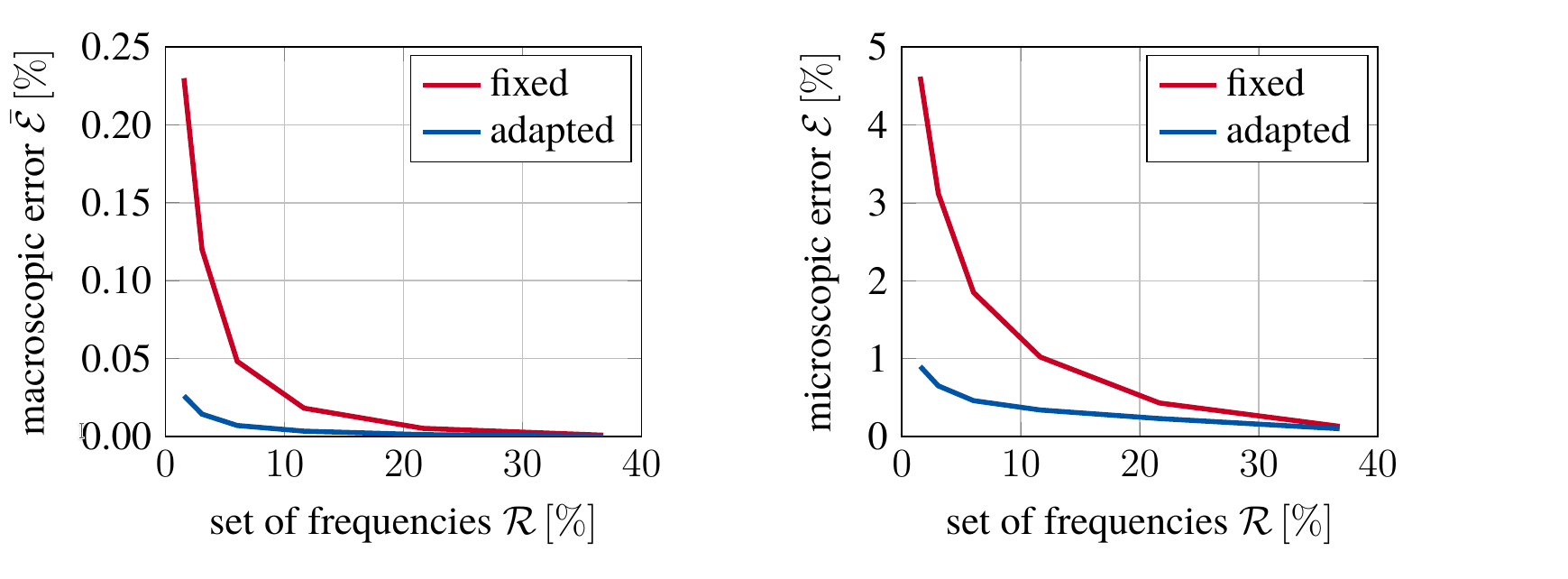}
	\caption{Macroscopic error $\bar{\mathcal E}$ (\textit{left}) and microscopic error $\mathcal E$ (\textit{right}) for the 2D elastic microstructure with one circular inclusion depending on the percentage of used frequencies $\mathcal R$ for the solution with the fixed and adapted sampling pattern.}
	\label{fig:OneIncError}
\end{figure}

To illustrate the adaptivity of the newly introduced sampling pattern generation and the resulting microstructural fields compared to the results with the fixed sampling pattern, a two dimensional microstructure with different centered inclusions as shown in Figure \ref{fig:OneInc} is investigated. Both, the inclusions and the matrix material are considerd to be elastic with $\lambda_\mathrm{I}=2.0 \, \text{GPa}$ and $\mu_\mathrm{I}=2.0 \, \text{GPa}$ for the inclusion and $\lambda_\mathrm{M}=1.0 \, \text{GPa}$ and $\mu_\mathrm{M}=1.0 \, \text{GPa}$ for the matrix material, respectively. The applied macroscopic strain is set to $\bar \varepsilon_{11}=0.01$ and $\bar \varepsilon_{22}=-0.01$.
First, we investigate the microstructure with one circular inclusion. Note, that the following observations have been made for all stress and strain fields. The results related to the fixed and adapted sampling pattern are shown in Figure \ref{fig:OneCircFixed} and Figure \ref{fig:OneCircAdapted}, respectively. In the top row, the sampling patterns for two reduced sets of frequencies ($\mathcal R = 0.78\,\%$ and $\mathcal R = 1.54\,\%$) are given. Subsequently, the corresponding stress fields $\sigma_{11}$ and the reference solution incorporating the full set of frequencies are shown. In order to better recognize the difference between the results for the reduced sets of frequencies compared to the reference solution, the bottom row shows the point-wise absolute difference in the microstructural stress fields $\Delta\sigma_{11}=|\sigma_{11}^\text{ref}-\sigma_{11}|$. 
As stated by \cite{Kochmann19} considering a fixed sampling pattern, the radial sampling pattern leads to good results in general since it considers a high amount of low frequencies, but also a certain amount of high frequencies. The low frequencies are necessary to capture the essential features of the microstructure, while the high frequencies are needed to capture for example microstructures with needle-like inclusions. Nevertheless, the radial sampling pattern leads to incoherent artifacts in the reduced solution, but which are needed in terms of the subsequent reconstruction algorithm. The difference in the microstructural stress fields compared to the reference solution is plotted in the last row of Figure \ref{fig:OneCircFixed}, while no reconstruction is incorporated up to now. It can be seen, that this error goes up to $0.01\,\text{GPa}$ in the area close to the edge of the inclusion, which is about one third compared to the maximum stress in the reference solution.\\
Choosing the reduced set of frequencies based on the geometrically adapted sampling pattern and compared to the fixed sampling pattern, a higher amount of low frequencies is needed to capture the microstructural geometry (see Figure \ref{fig:OneCircAdapted}). Using this reduced set of frequencies also in terms of the microstructure simulation, more accurate results are generated. The highest error occurs in the transition from matrix to inclusion, which arises from the very high amount of frequencies which are necessary to capture such a sharp transition.
The macroscopic error $\bar{\mathcal E}$ and the microscopic error $\mathcal{E}$ are shown in Figure \ref{fig:OneIncError} depending on the reduced set of frequencies $\mathcal R$. Since a simple microstructure with only one centered circular inclusion is investigated in this example, the macroscopic (\textit{left}) and the microscopic errors (\textit{right}) are in both cases quite low. Nevertheless, the error corresponding to the adapted sampling pattern is always significantly smaller - especially, when the reduced set of frequencies consists only of few frequencies. For a higher number of frequencies, the errors of both solutions converge towards zero. 

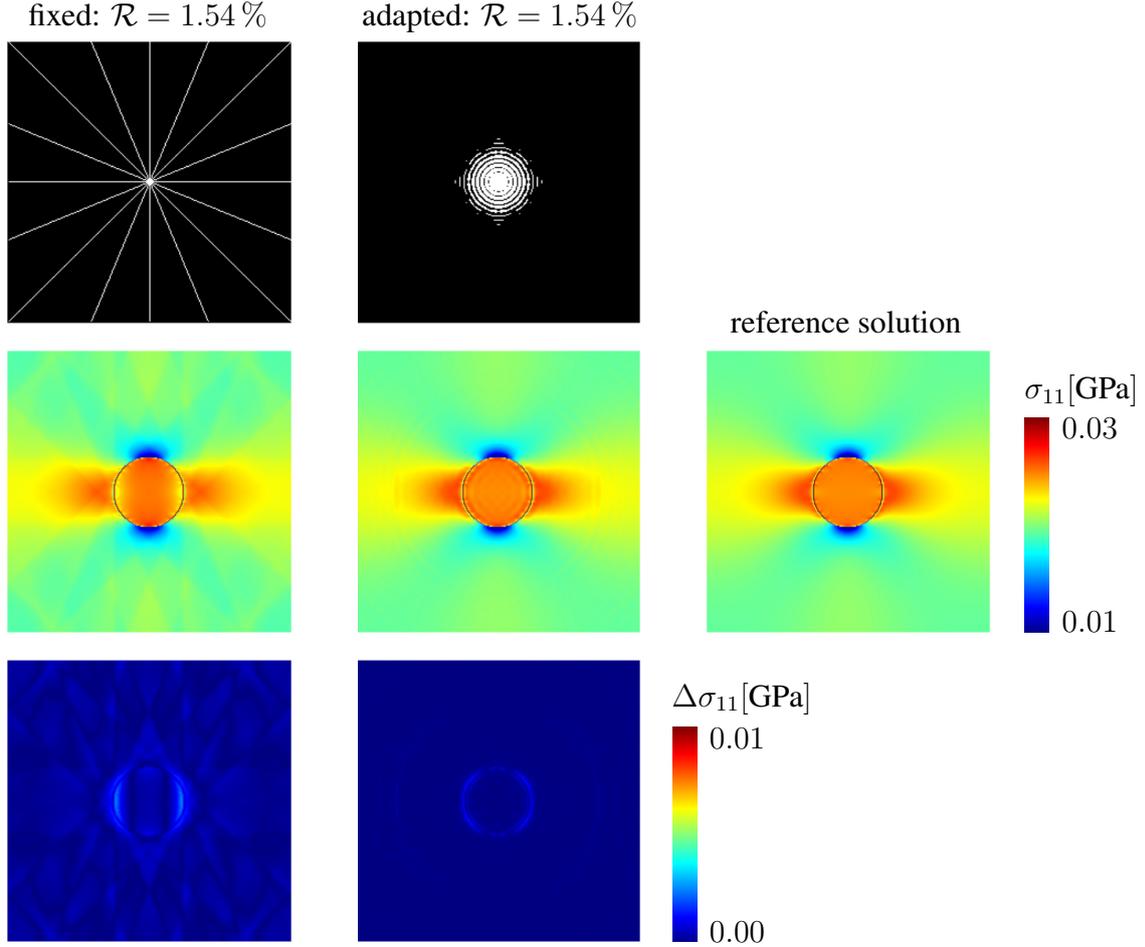
\begin{figure}[H]
	\hspace{-1.5cm}
	\centering
	\includegraphics[width=0.95\textwidth]{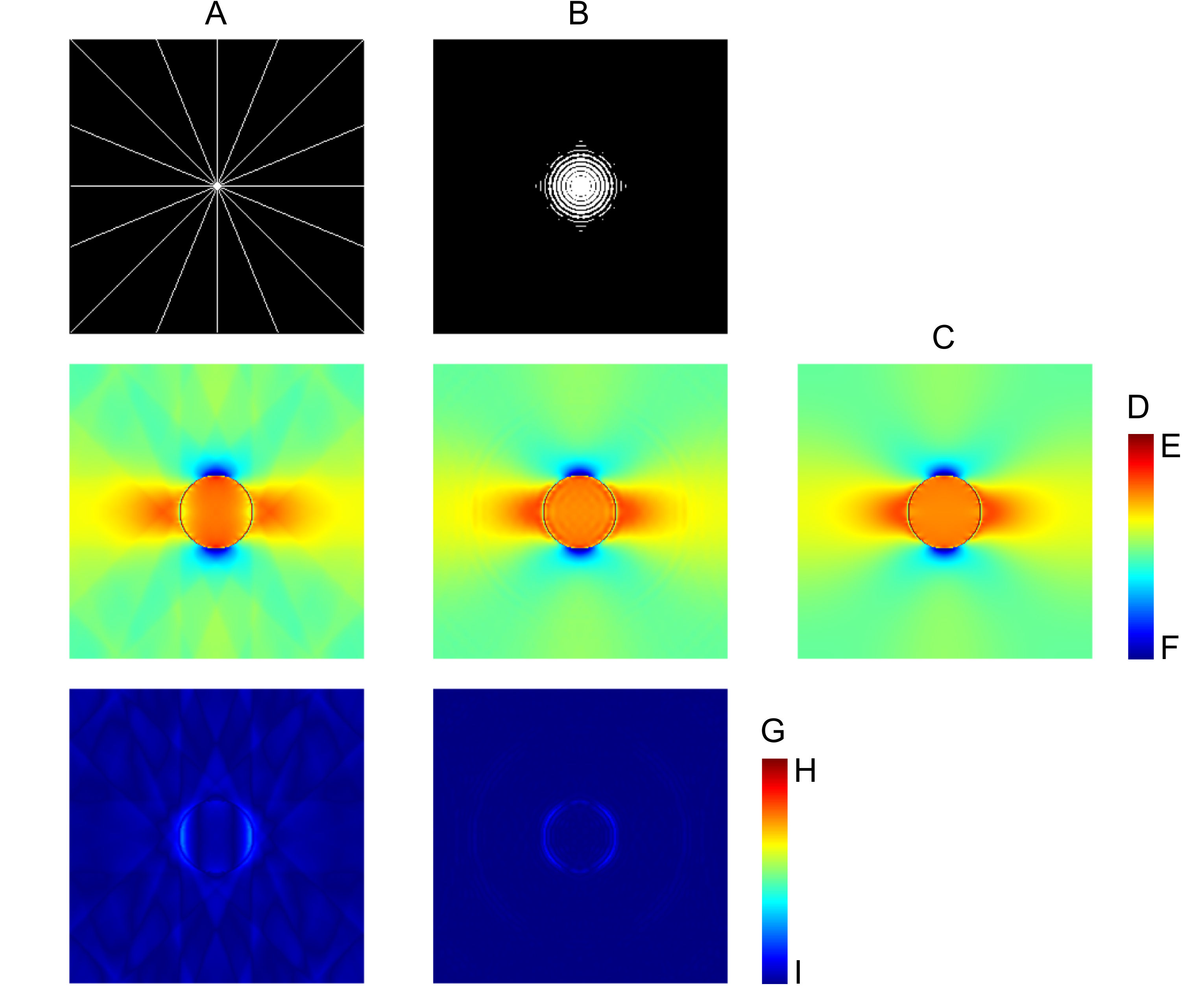}
	\put(-403,351){\colorbox{white}{fixed: $\mathcal R = 1.54\,\%$}}
	\put(-278,351){\colorbox{white}{adapted: $\mathcal R = 1.54\,\%$}}
	\put(-140,235){\colorbox{white}{reference solution}}
	\put(-30,210){\colorbox{white}{$\sigma_{11}[\text{GPa}]$}}
	\put(-16,195){\colorbox{white}{$0.03$}}
	\put(-16,122){\colorbox{white}{$0.01$}}
	\put(-162,94){\colorbox{white}{$\Delta\sigma_{11}[\text{GPa}]$}}
	\put(-148,78){\colorbox{white}{$0.01$}}
	\put(-148,5){\colorbox{white}{$0.00$}}
	\caption{Microstructural fields for the 2D elastic microstructure with one circular inclusion. \textit{Top row}: Fixed and adapted sampling pattern with the same number of wave vectors. \textit{Middle row}: Corresponding microstructural stress field $\sigma_{11}$ incorporating the reconstruction and compatibility step for the solution of the fixed sampling pattern and only the compatibility step for the solution of the adapted sampling pattern and reference stress field computed with the full set of frequencies. \textit{Bottom row}: Absolute difference in the microstructural stress field $\Delta\sigma_{11}$.}
	\label{fig:OneIncCircComp}
\end{figure}

\begin{figure}[H]
	\includegraphics[width=\textwidth]{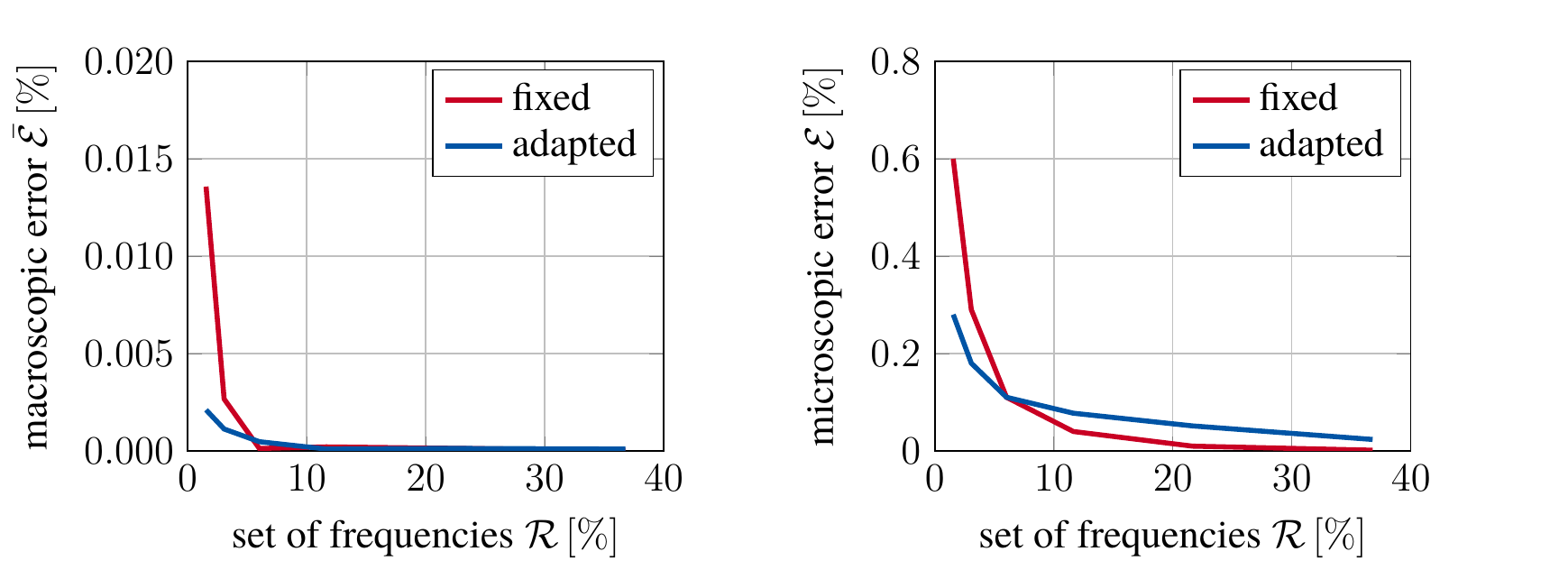}
	\caption{Macroscopic error $\bar{\mathcal E}$ (left) and microscopic error $\mathcal E$ (right) for the 2D elastic microstructure with one circular inclusion depending on the reduced set of frequencies $\mathcal R$ for the solution with the fixed sampling pattern with reconstruction and compatibility step and for the adapted sampling pattern only with the compatibility step.}
	\label{fig:OneIncErrorComp}
\end{figure}

Using the reconstruction algorithm and the compatibility step the solution for the fixed sampling pattern is improved as shown in Figure \ref{fig:OneIncCircComp} in the left column. Since the microstructural fields related to the adapted sampling pattern do not have incoherent artifacts, the reconstruction algorithm does not yield any further improvement of the result and is therefore not needed. Instead, just the so-called compatibility step is applied. Thus, the Lippmann-Schwinger equation is solved once using the full set of frequencies based on the stress and the strain fields from the reduced simulation. It should be mentioned that in this context the earlier given name might be misleading, since the solutions are already compatible. The corresponding solutions are shown in Figure \ref{fig:OneIncCircComp} in the centered column. It can be seen, that the micromechanical solution fields which are related to the adapted sampling pattern are still better compared to the solutions based on the fixed sampling pattern after the reconstruction.    
Incorporating this post-processing step, the macroscopic error $\bar{\mathcal E}$ and microscopic error $\mathcal{E}$ are shown in Figure \ref{fig:OneIncErrorComp}. For the fixed sampling pattern, the error refers to the solution after the reconstruction and the compatibility step. For the adapted sampling pattern, the error corresponds to the compatibility step, only. For both cases, Figure \ref{fig:OneIncErrorComp} shows the error depending on the percentage of used frequencies $\mathcal R$. The difference between the fixed and the adapted sampling pattern is rather small, while the error of the adapted sampling pattern is smaller for $\mathcal R < 6\,\%$ but larger for $\mathcal R \ge 6\,\%$. The range $\mathcal R < 6\,\%$ is of much larger interest, since the highest speed-up is gained by a set with a low number of frequencies.

\begin{figure}[h]
	\hspace{-1.5cm}
	\centering
	\includegraphics[width=0.95\textwidth]{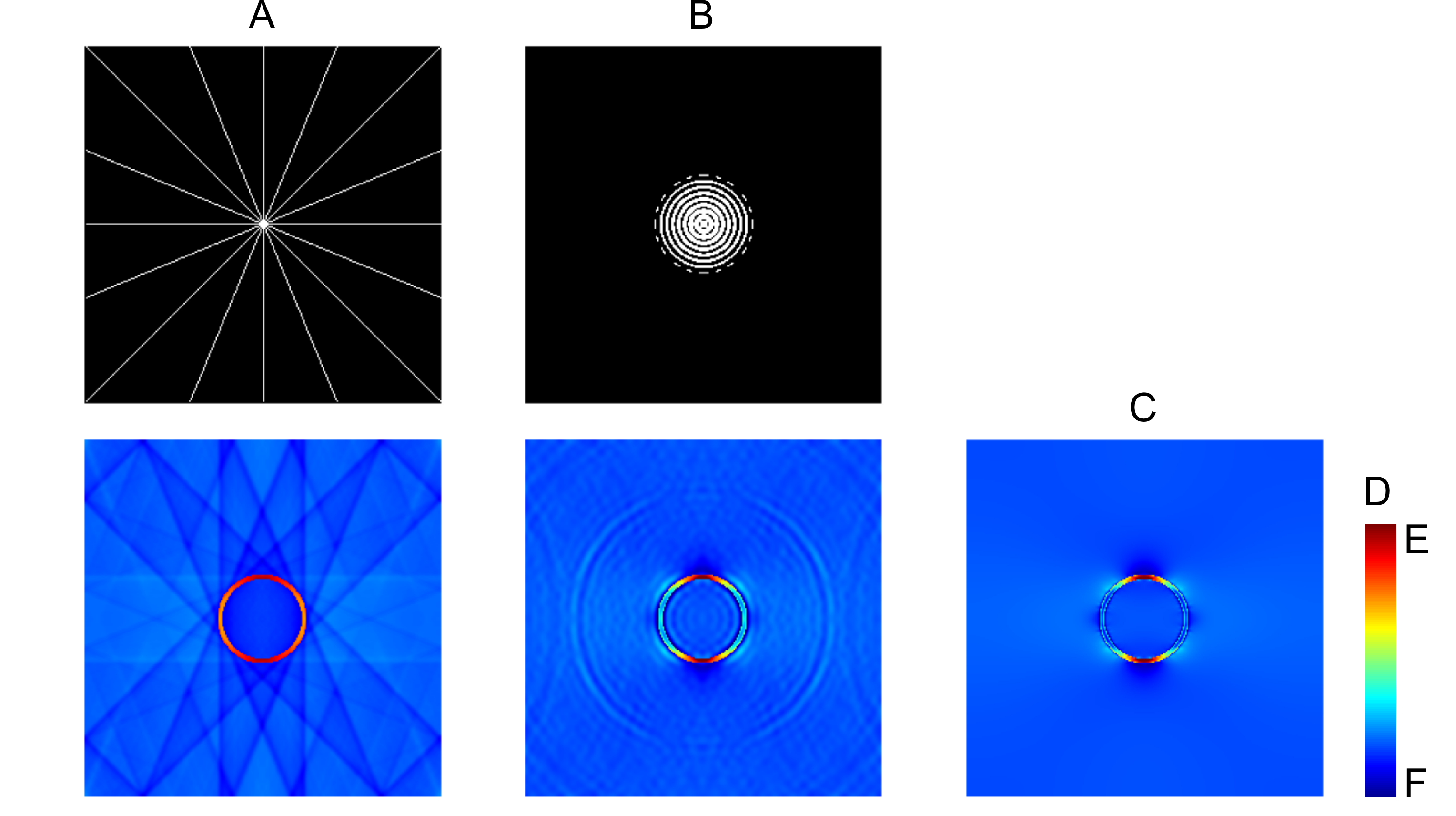}
	\put(-402,235){\colorbox{white}{fixed: $\mathcal R = 1.54\,\%$}}
	\put(-278,235){\colorbox{white}{adapted: $\mathcal R = 1.54\,\%$}}
	\put(-140,119){\colorbox{white}{reference solution}}
	\put(-30,95){\colorbox{white}{$\sigma_{11}[\text{GPa}]$}}
	\put(-16,79){\colorbox{white}{$0.040$}}
	\put(-16,6){\colorbox{white}{$0.015$}}
	\caption{Microstructural fields for the 2D elastic microstructure with one annular inclusion. \textit{Top row}: Fixed and adapted sampling pattern with the same number of wave vectors. \textit{Bottom row}: Corresponding microstructural stress field $\sigma_{11}$ and reference stress field computed with the full set of frequencies.}
	\label{fig:OneIncRing}
\end{figure}

Tables \ref{tab:OneIncTimeFixed} and \ref{tab:OneIncTimeAdapted} present the total CPU times for the computations with the fixed and adapted sampling pattern, respectively. These CPU times are subdivided into the mean CPU time per iteration step for solving the convolution integral and the mean CPU time per iteration step for solving the constitutive law. The time for solving the constitutive law is almost independent of the number of considered frequencies. The speed up in the total CPU times is gained by solving the convolution integral in Fourier space with the reduced set of frequencies. Independent of the sampling pattern, the speed up factor for $\mathcal R = 1.54\,\%$ of frequencies is about $7$-$8$, while the results with the adapted sampling pattern are more accurate. In addition, Table \ref{tab:OneIncTimeFixed} shows the CPU times for the time consuming reconstruction step, which is not necessary for the adapted sampling pattern and the CPU time for the compatibility step, which is almost the same for the fixed and adapted sampling pattern (see Table \ref{tab:OneIncTimeAdapted}).
\begin{table}[H]
	\centering
	\begin{tabular}{|c||c|c|c||c||c|}
		\hline
		elastic & \multicolumn{5}{c|}{\textbf{CPU time [s] - fixed sampling pattern}}\\ \hline \hline
		\(\mathcal{R}\,[\%]\) & total & \(-\hat{\BbbGammaVar}^{(0)}\hat{\bm{\tau}}(\bm{\varepsilon})\) (mean) & \(\bm{\sigma}(\bm{\varepsilon})\) (mean) & reconstruction & compatibility \\ \hline
		1.54   & 0.448 & 0.015 & 0.006 & 58.295 & 0.324  \\ \hline
		3.06   & 0.558 & 0.022 & 0.006 & 54.679 & 0.333  \\ \hline
		6.02   & 0.759 & 0.036 & 0.005 & 56.764 & 0.337  \\ \hline
		11.64  & 0.958 & 0.061 & 0.005 & 59.208 & 0.354  \\ \hline
		21.66  & 1.253 & 0.106 & 0.005 & 47.910 & 0.341  \\ \hline
		36.79  & 1.869 & 0.166 & 0.005 & 49.136 & 0.327  \\ \hline 
		\vdots & \vdots & \vdots & \vdots & \vdots & \vdots \\ \hline
		unreduced& 3.196  & 0.302 & 0.004 & - & -\\ \hline
	\end{tabular}
	\caption{Total CPU time with mean CPU time per iteration step for solving the convolution integral and the constitutive law and CPU times for the reconstruction and the compatibility step of the simulation with the fixed sampling pattern for the 2D elastic microstructure with one circular inclusion.}
	\label{tab:OneIncTimeFixed}
\end{table}
\begin{table}[H]
	\centering
	\begin{tabular}{|c||c|c|c||c||c|}
		\hline
		elastic & \multicolumn{5}{c|}{\textbf{CPU time [s] - adapted sampling pattern}}\\ \hline \hline
		\(\mathcal{R}\,[\%]\) & & total \(-\hat{\BbbGammaVar}^{(0)}\hat{\bm{\tau}}(\bm{\varepsilon})\) (mean) & \(\bm{\sigma}(\bm{\varepsilon})\) (mean) & reconstruction & compatibility\\ \hline
		1.54  & 0.388 & 0.013 & 0.005  & - & 0.341 \\ \hline
		3.06  & 0.495 & 0.022 & 0.005  & - & 0.378 \\ \hline
		6.02  & 0.641 & 0.034 & 0.005  & - & 0.346 \\ \hline
		11.64 & 0.948 & 0.060 & 0.005  & - & 0.322 \\ \hline
		21.66 & 1.412 & 0.099 & 0.005  & - & 0.340 \\ \hline
		36.79 & 1.912 & 0.169 & 0.005  & - & 0.350 \\ \hline 
		\vdots & \vdots & \vdots & \vdots & \vdots & \vdots \\ \hline
		unreduced & 3.196 & 0.302 & 0.004 & - & - \\ \hline
	\end{tabular}
	\caption{Total CPU time with mean CPU time per iteration step for solving the convolution integral and the constitutive law and CPU times for the reconstruction and the compatibility step of the simulation with the adapted sampling pattern for the 2D elastic microstructure with one circular inclusion.}
	\label{tab:OneIncTimeAdapted}
\end{table}
The same investigations could be made using the microstructures with the annular, the elliptical, and the quadratic inclusion shown in Figure \ref{fig:OneInc} and would lead to similar results. Due to that, we just show the sampling patterns and the corresponding microstructural stress fields $\sigma_{11}$ for $\mathcal R = 1.54\,\%$ of frequencies considering these microstructures in Figures \ref{fig:OneIncRing} - \ref{fig:OneIncSquare}. The results show the relation between the arrangement of the considered frequencies in the adapted sampling pattern and the geometry of the inclusion within the matrix. Additionally it can be seen, that the results gained by the new approach are always closer to the reference solution than the solutions based on the fixed sampling pattern.\\
In Figure \ref{fig:OneIncRing}, the geometrically adapted sampling pattern of the microstructure with an annular inclusion is shown. This sampling pattern is similar to the adapted sampling corresponding to the circular inclusion; see Figure \ref{fig:OneCircAdapted}. Nevertheless, the amount of high frequencies is slightly higher. The adapted sampling pattern corresponding to the elliptical inclusion, shown in Figure \ref{fig:OneIncEllipse}, differs totally from that. It can be seen, that in direction of the major axis of the ellipse lower frequencies are needed and perpendicular to that higher frequencies are necessary, since a smaller distance needs to be bridged in this direction. As a last example, Figure \ref{fig:OneIncSquare} shows the microstructure with a quadratic inclusion. The sharp edges of this last examined type of inclusion again lead to a totally different set of frequencies.
\begin{figure}[H]
	\hspace{-1.5cm}
	\centering
	\includegraphics[width=0.95\textwidth]{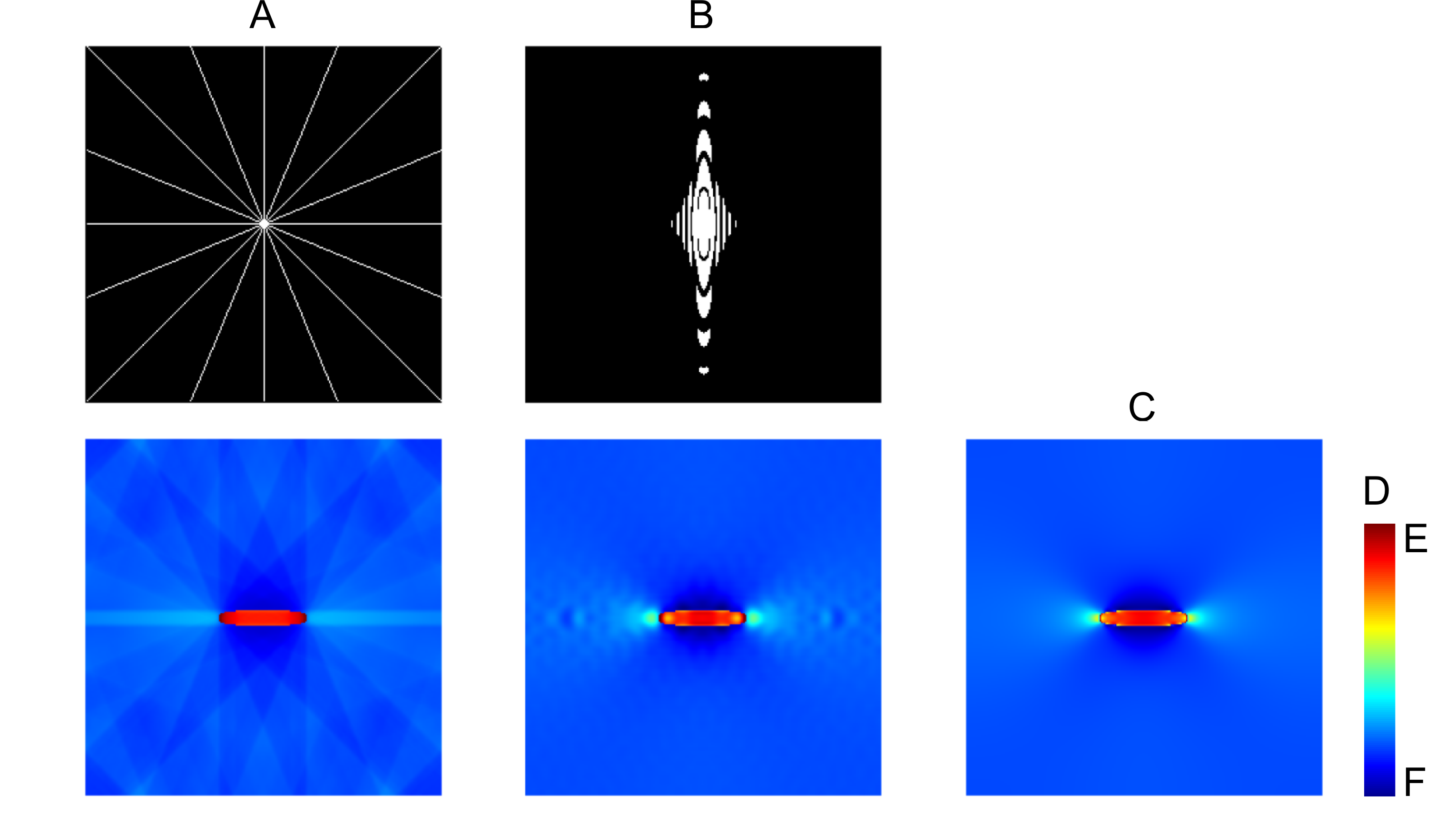}
	\put(-402,235){\colorbox{white}{fixed: $\mathcal R = 1.54\,\%$}}
	\put(-278,235){\colorbox{white}{adapted: $\mathcal R = 1.54\,\%$}}
	\put(-140,119){\colorbox{white}{reference solution}}
	\put(-30,95){\colorbox{white}{$\sigma_{11}[\text{GPa}]$}}
	\put(-16,79){\colorbox{white}{$0.040$}}
	\put(-16,6){\colorbox{white}{$0.015$}}
	\caption{Microstructural fields for the 2D elastic microstructure with one elliptical inclusion. \textit{Top row}: Fixed and adapted sampling pattern with the same number of wave vectors. \textit{Bottom row}: Corresponding microstructural stress field $\sigma_{11}$ and reference stress field computed with the full set of frequencies.}
	\label{fig:OneIncEllipse}
\end{figure}
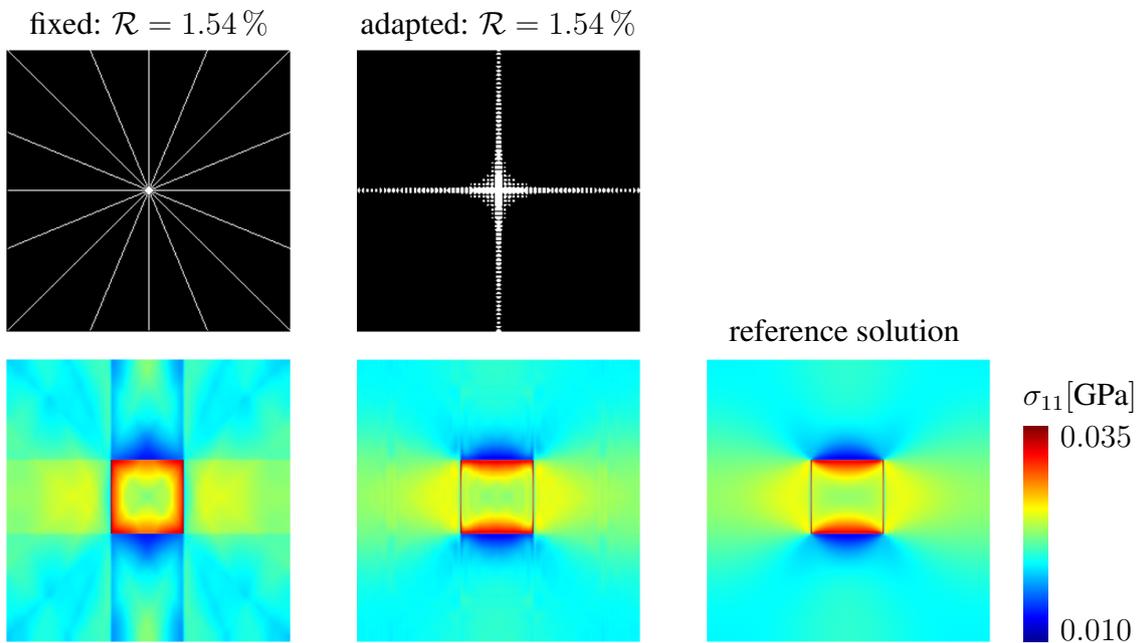
\begin{figure}[H]l
	\hspace{-1.5cm}
	\centering
	\includegraphics[width=0.95\textwidth]{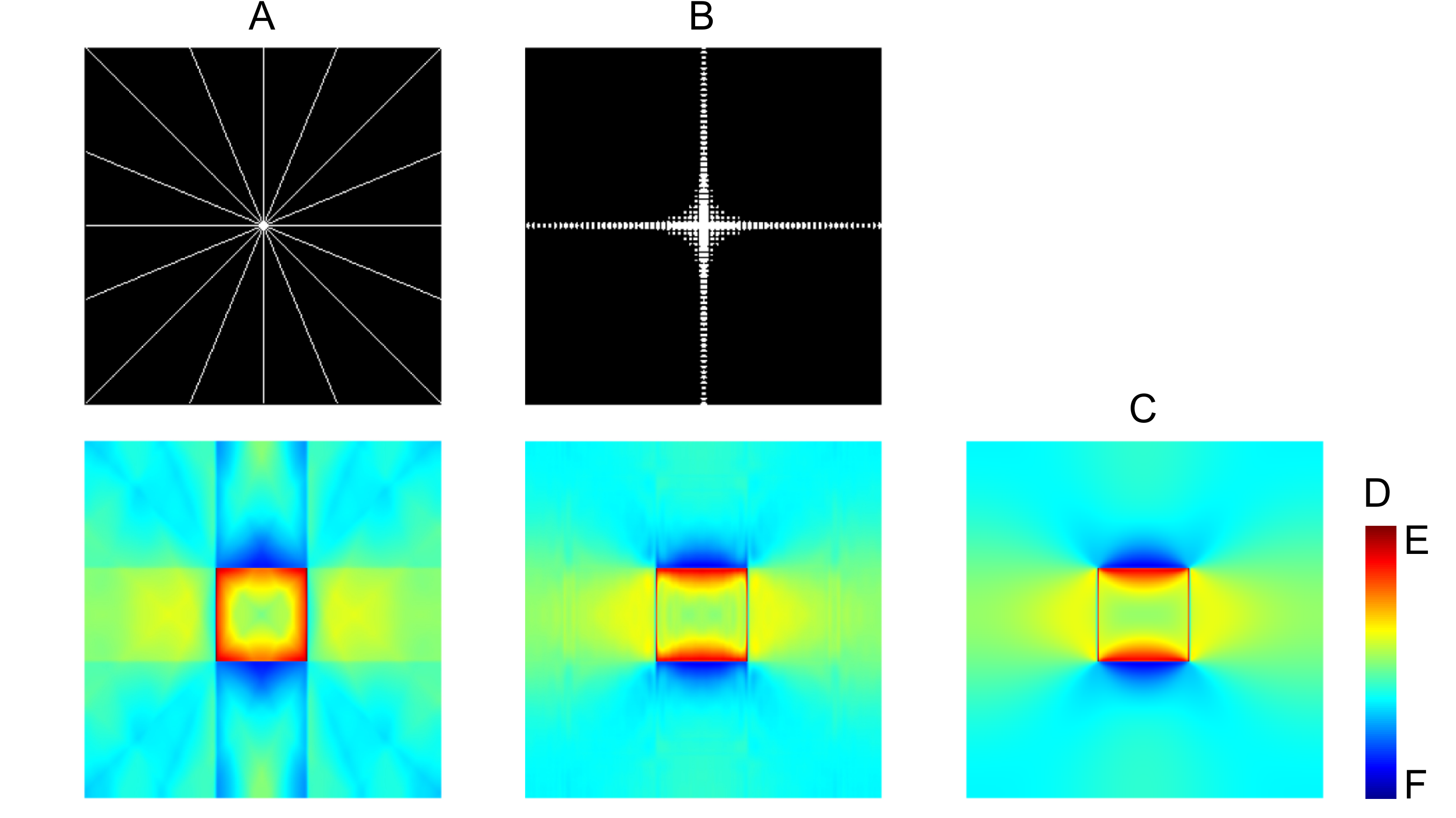}
	\put(-402,235){\colorbox{white}{fixed: $\mathcal R = 1.54\,\%$}}
	\put(-278,235){\colorbox{white}{adapted: $\mathcal R = 1.54\,\%$}}
	\put(-140,119){\colorbox{white}{reference solution}}
	\put(-30,95){\colorbox{white}{$\sigma_{11}[\text{GPa}]$}}
	\put(-16,79){\colorbox{white}{$0.035$}}
	\put(-16,6){\colorbox{white}{$0.010$}}
	\caption{Microstructural fields for the 2D elastic microstructure with one quadratic inclusion. \textit{Top row}: Fixed and adapted sampling pattern with the same number of wave vectors. \textit{Bottom row}: Corresponding microstructural stress field $\sigma_{11}$ and reference stress field computed with the full set of frequencies.}	
	\label{fig:OneIncSquare}
\end{figure}

\subsection{Elastic and elasto-plastic 2D two phase material with several inclusions}
\label{chap:several}
\begin{figure}[H]
	\begin{minipage}{0.5\textwidth}
		\centering
		\includegraphics[width=0.7\textwidth]{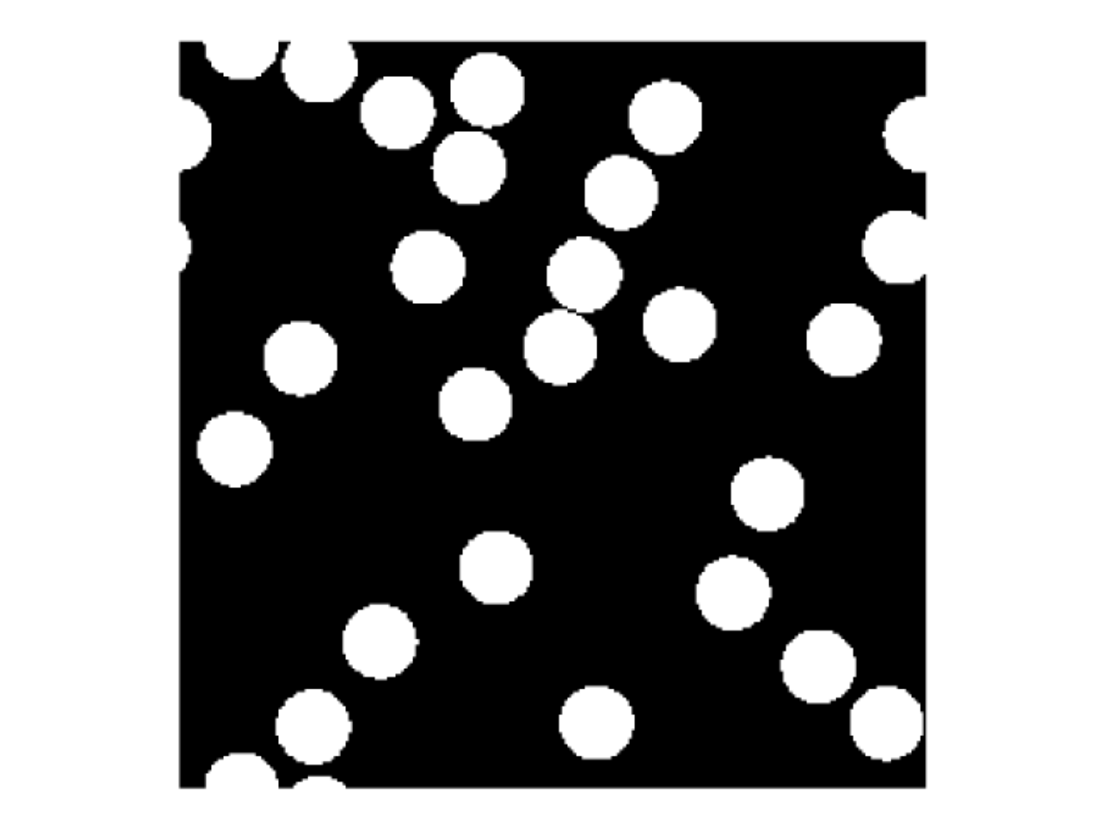}
	\end{minipage}%
	\begin{minipage}{0.5\textwidth}
		\begin{eqnarray}
		\nonumber
		\bar{\boldsymbol\varepsilon}=
		\begin{pmatrix}
		0.01 & 0.002 \\
		0.002 & -0.01
		\end{pmatrix}
		\end{eqnarray}
	\end{minipage}
	\caption{Composite with several circular inclusions and prescribed macroscopic stain.}
	\label{fig:MicroSeveral}
\end{figure}

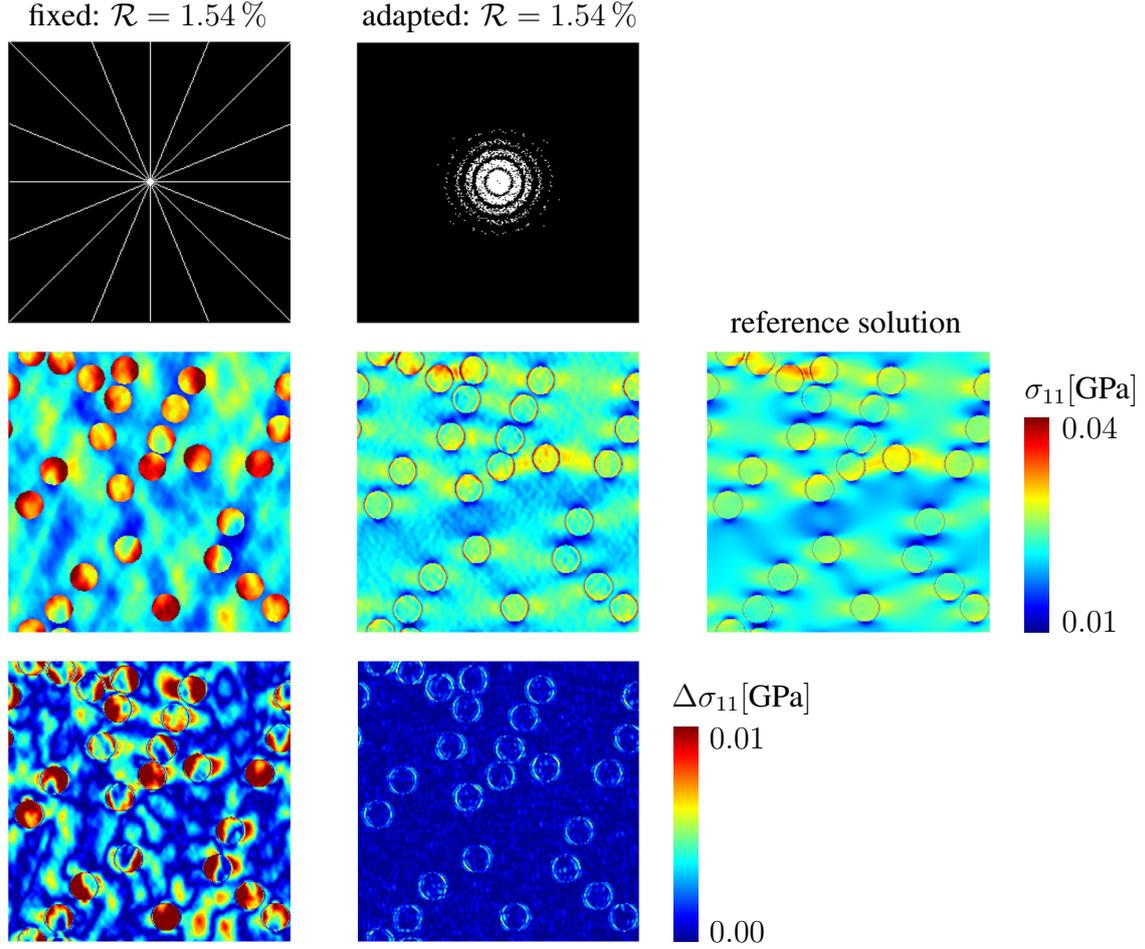
\begin{figure}[H]
	\hspace{-1.5cm}
	\centering
	\includegraphics[width=0.95\textwidth]{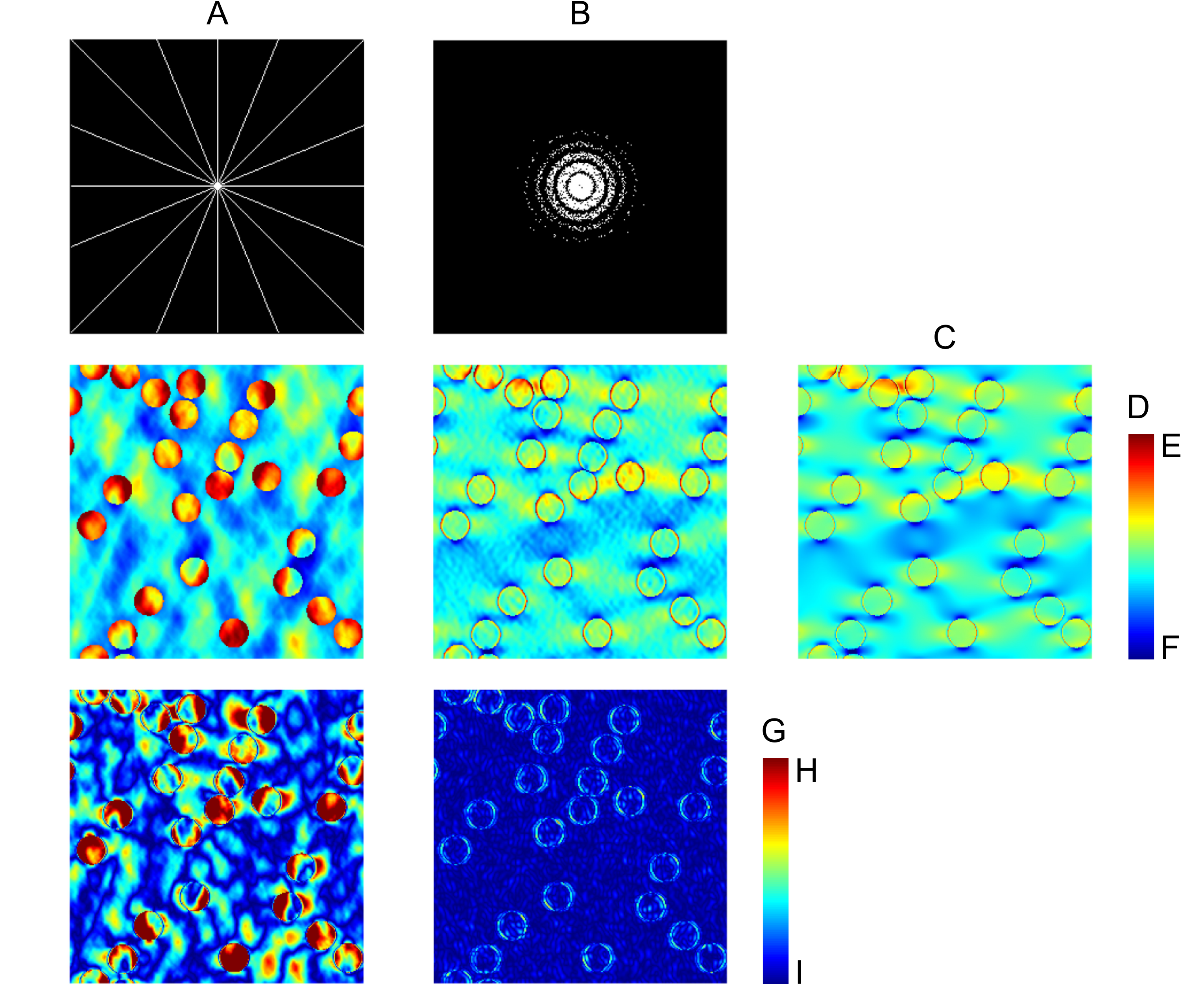}
	\put(-403,351){\colorbox{white}{fixed: $\mathcal R = 1.54\,\%$}}
	\put(-278,351){\colorbox{white}{adapted: $\mathcal R = 1.54\,\%$}}
	\put(-140,235){\colorbox{white}{reference solution}}
	\put(-30,210){\colorbox{white}{$\sigma_{11}[\text{GPa}]$}}
	\put(-16,195){\colorbox{white}{$0.04$}}
	\put(-16,122){\colorbox{white}{$0.01$}}
	\put(-162,94){\colorbox{white}{$\Delta\sigma_{11}[\text{GPa}]$}}
	\put(-148,78){\colorbox{white}{$0.01$}}
	\put(-148,5){\colorbox{white}{$0.00$}}
	\caption{Microstructural fields for the 2D elastic microstructure with several circular inclusions. \textit{Top row}: Fixed and adapted sampling pattern with the same number of wave vectors. \textit{Middle row}: Corresponding microstructural stress field $\sigma_{11}$ and reference stress field computed with the full set of frequencies. \textit{Bottom row}: Absolute difference in the microstructural stress field $\Delta\sigma_{11}$.}
	\label{fig:SeveralRedu}
\end{figure}

\begin{figure}[H]
	\hspace{-1.5cm}
	\centering
	\includegraphics[width=0.95\textwidth]{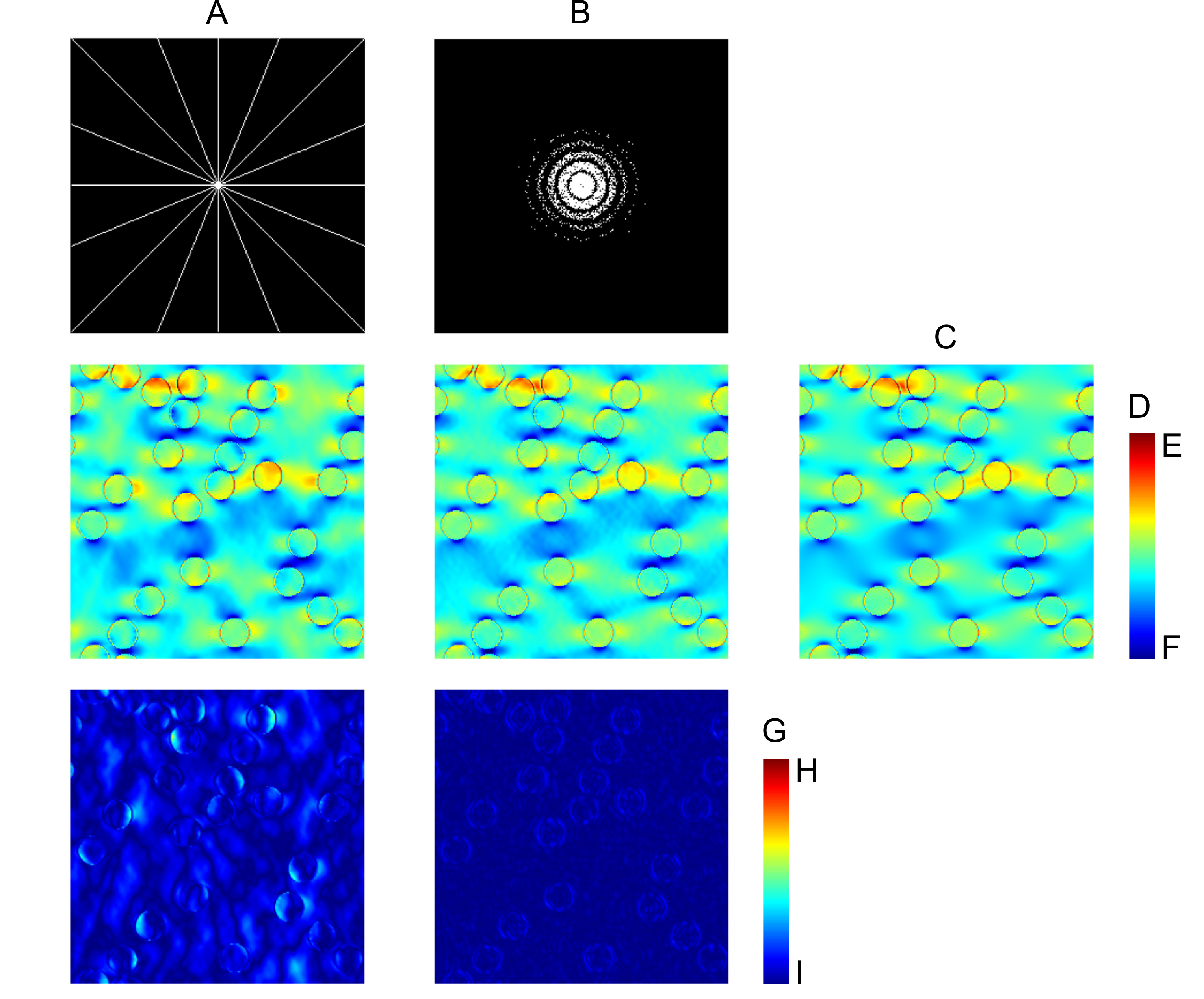}
	\put(-403,351){\colorbox{white}{fixed: $\mathcal R = 1.54\,\%$}}
	\put(-278,351){\colorbox{white}{adapted: $\mathcal R = 1.54\,\%$}}
	\put(-140,235){\colorbox{white}{reference solution}}
	\put(-30,210){\colorbox{white}{$\sigma_{11}[\text{GPa}]$}}
	\put(-16,195){\colorbox{white}{$0.04$}}
	\put(-16,122){\colorbox{white}{$0.01$}}
	\put(-162,94){\colorbox{white}{$\Delta\sigma_{11}[\text{GPa}]$}}
	\put(-148,78){\colorbox{white}{$0.01$}}
	\put(-148,5){\colorbox{white}{$0.00$}}
	\caption{Microstructural fields for the 2D elastic microstructure with several circular inclusions. \textit{Top row}: Fixed and adapted sampling pattern with the same number of wave vectors. \textit{Middle row}: Corresponding microstructural stress field $\sigma_{11}$ incorporating the reconstruction and compatibility step for the solution of the fixed sampling pattern and only the compatibility step for the solution of the adapted sampling pattern and reference stress field computed with the full set of frequencies. \textit{Bottom row}: Absolute difference in the microstructural stress field $\Delta\sigma_{11}$.}
	\label{fig:SeveralComp}
\end{figure}
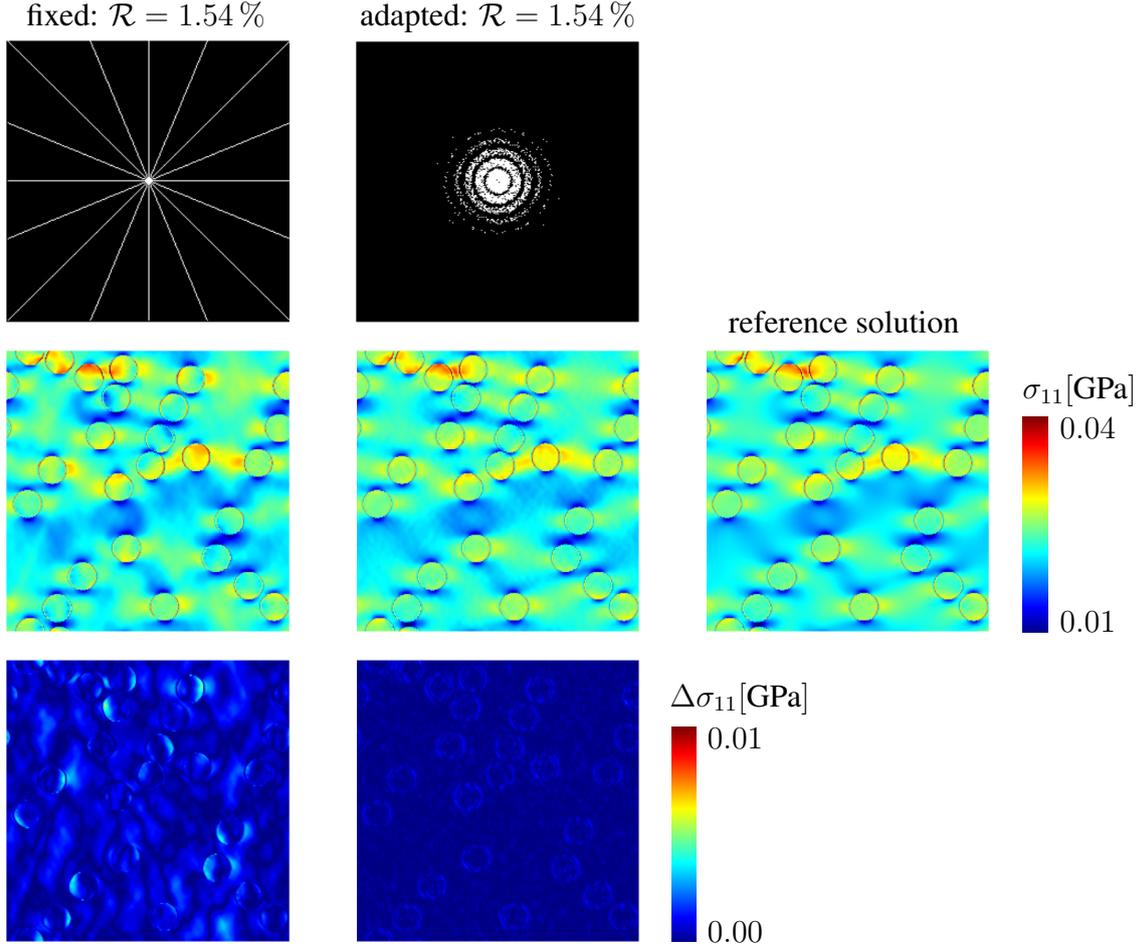

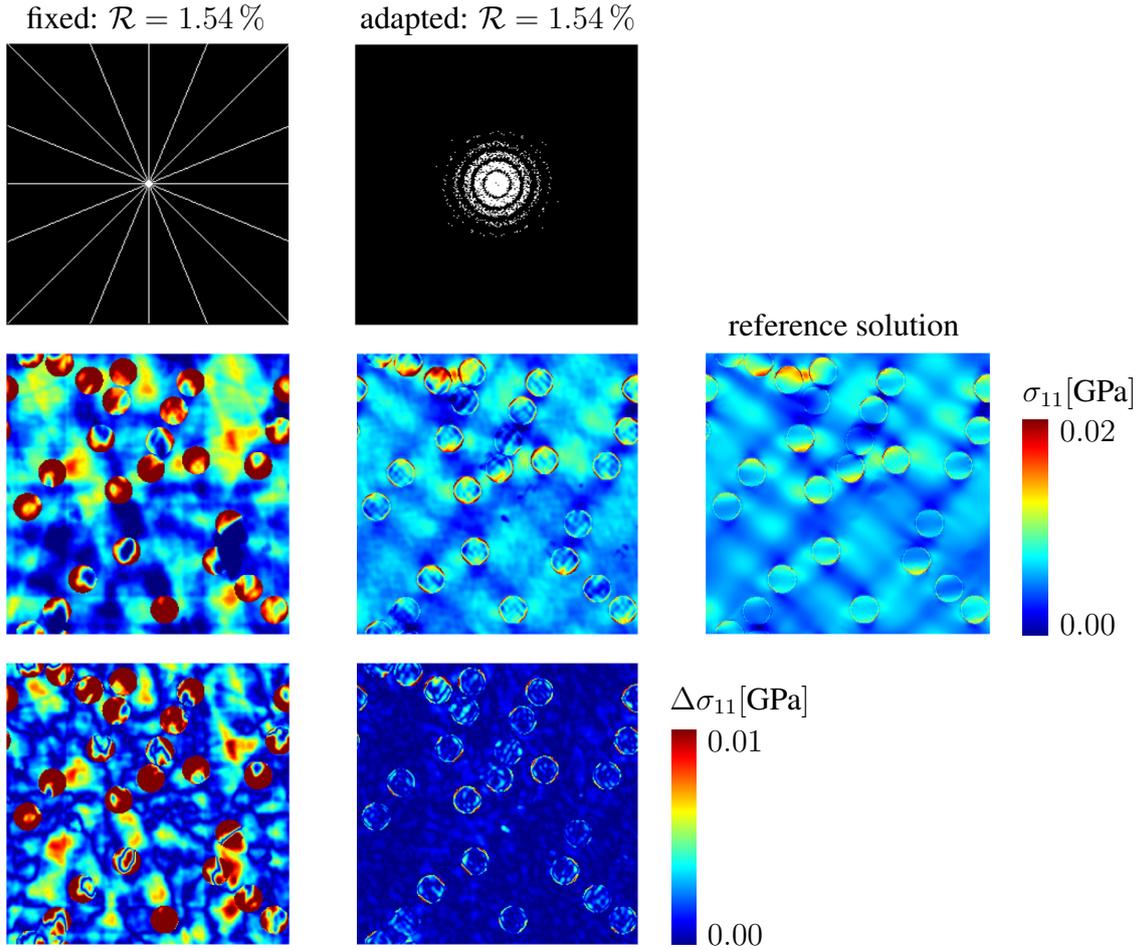
\begin{figure}[H]
	\hspace{-1.5cm}
	\centering
	\includegraphics[width=0.95\textwidth]{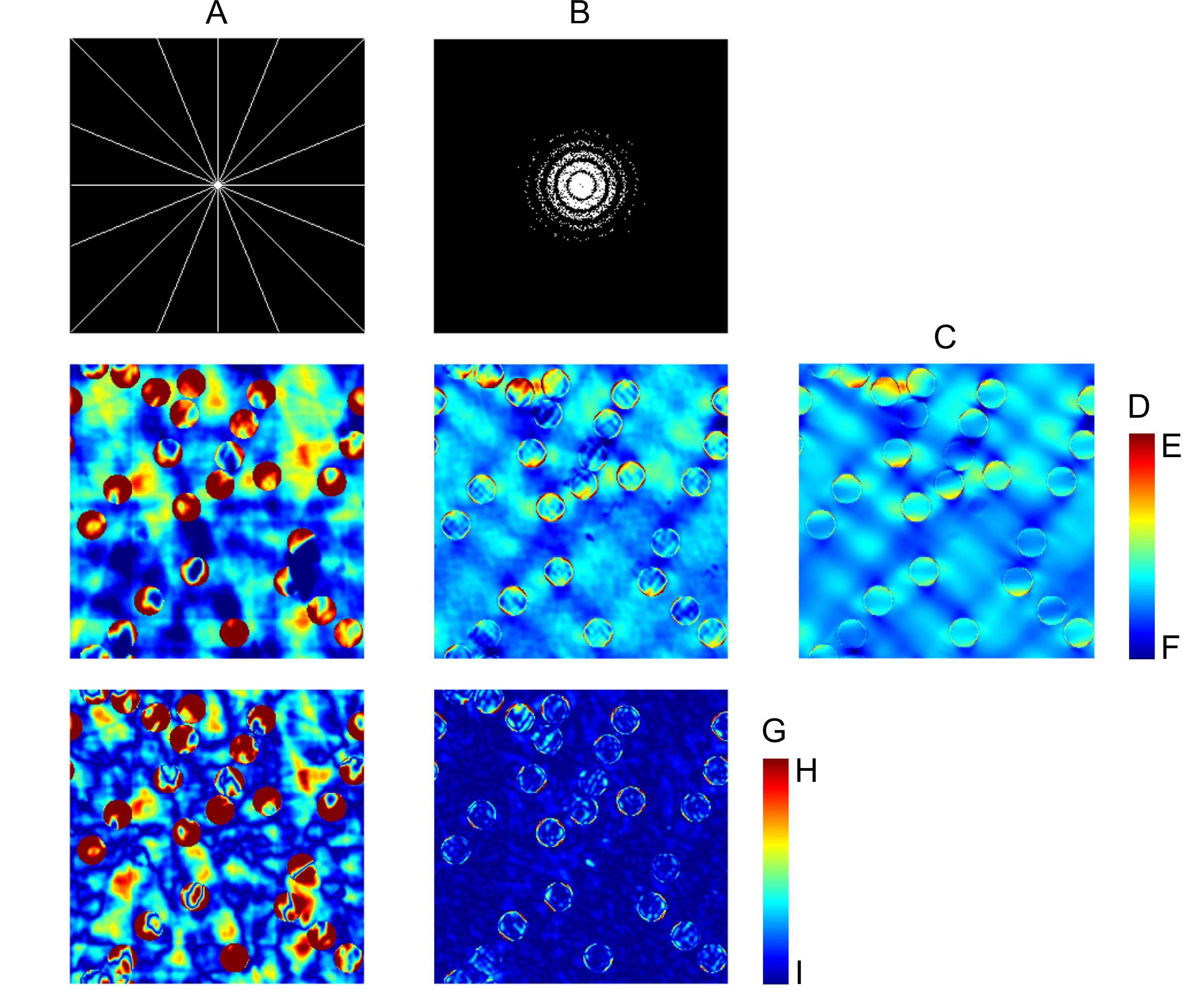}
	\put(-403,351){\colorbox{white}{fixed: $\mathcal R = 1.54\,\%$}}
	\put(-278,351){\colorbox{white}{adapted: $\mathcal R = 1.54\,\%$}}
	\put(-140,235){\colorbox{white}{reference solution}}
	\put(-30,210){\colorbox{white}{$\sigma_{11}[\text{GPa}]$}}
	\put(-16,195){\colorbox{white}{$0.02$}}
	\put(-16,122){\colorbox{white}{$0.00$}}
	\put(-162,94){\colorbox{white}{$\Delta\sigma_{11}[\text{GPa}]$}}
	\put(-148,78){\colorbox{white}{$0.01$}}
	\put(-148,5){\colorbox{white}{$0.00$}}
	\caption{Microstructural fields for the 2D elasto-plastic microstructure with several circular inclusions. \textit{Top row}: Fixed and adapted sampling pattern with the same number of wave vectors. \textit{Middle row}: Corresponding microstructural stress field $\sigma_{11}$ and reference stress field computed with the full set of frequencies. \textit{Bottom row}: Absolute difference in the microstructural stress field $\Delta\sigma_{11}$.}
	\label{fig:SeveralPlastRedu}
\end{figure}

\begin{figure}[H]
	\includegraphics[width=\textwidth]{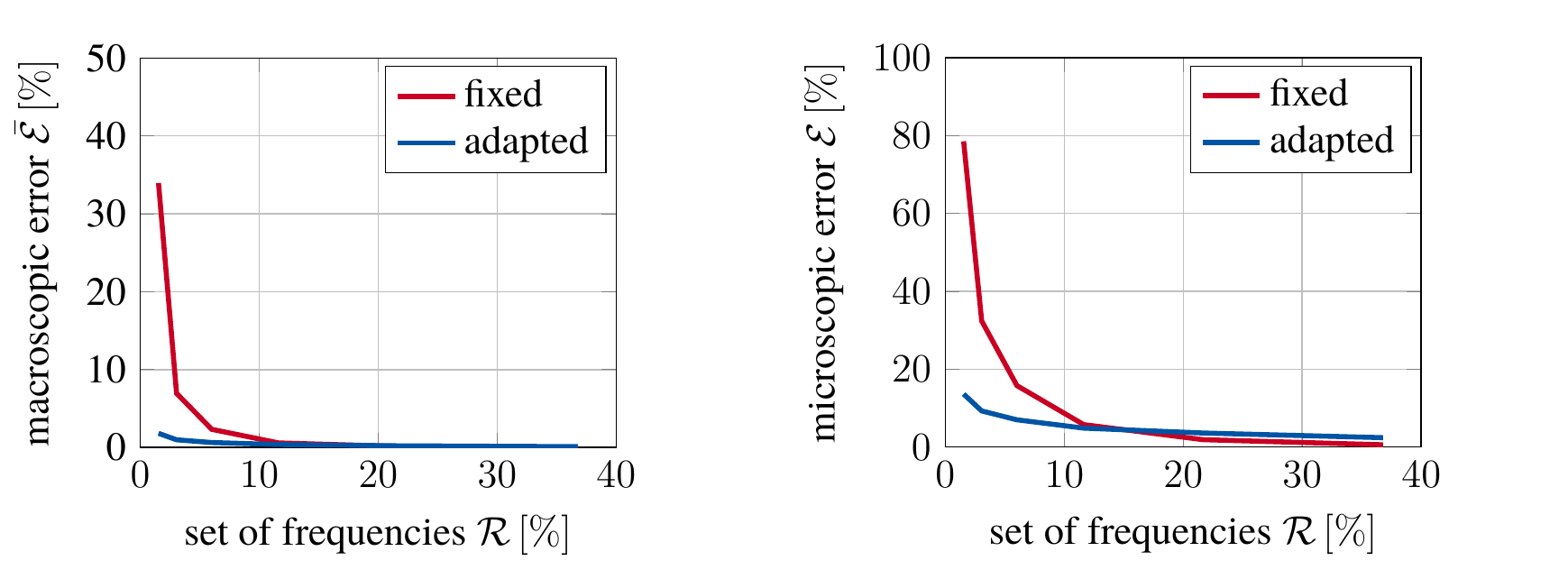}
	\caption{Macroscopic error $\bar{\mathcal E}$ (\textit{left}) and microscopic error $\mathcal E$ (\textit{right}) for the 2D elasto-plastic microstructure with several circular inclusions depending on the percentage of used frequencies $\mathcal R$ for the solution with the fixed and adapted sampling pattern.}
	\label{fig:ErrorPlast}
\end{figure}

\begin{figure}[H]
	\hspace{-1.5cm}
	\centering
	\includegraphics[width=0.95\textwidth]{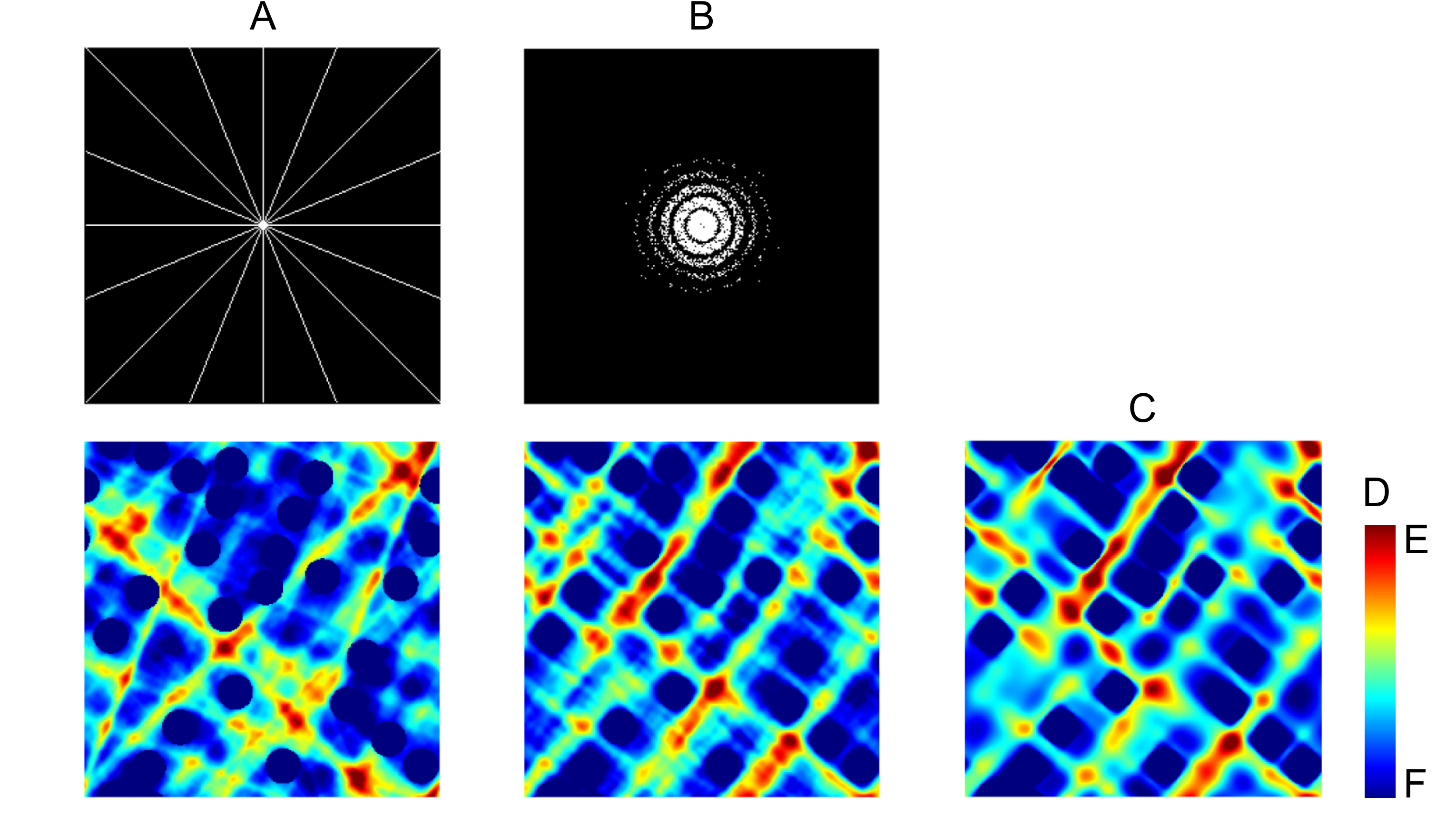}
	\put(-402,235){\colorbox{white}{fixed: $\mathcal R = 1.54\,\%$}}
	\put(-278,235){\colorbox{white}{adapted: $\mathcal R = 1.54\,\%$}}
	\put(-140,119){\colorbox{white}{reference solution}}
	\put(-30,95){\colorbox{white}{$\varepsilon_p^{acc}[\text{-}]$}}
	\put(-16,79){\colorbox{white}{$0.03$}}
	\put(-16,6){\colorbox{white}{$0.00$}}
	\caption{Microstructural fields for the 2D elasto-plastic microstructure with several circular inclusions. \textit{Top row}: Fixed and adapted sampling pattern with the same number of wave vectors. \textit{Bottom row}: Corresponding microstructural accumulated plastic strain field $\varepsilon_p^{acc}$ and reference accumulated plastic strain field computed with the full set of frequencies.}
	\label{fig:SeveralPlastAlphaRedu}
\end{figure}
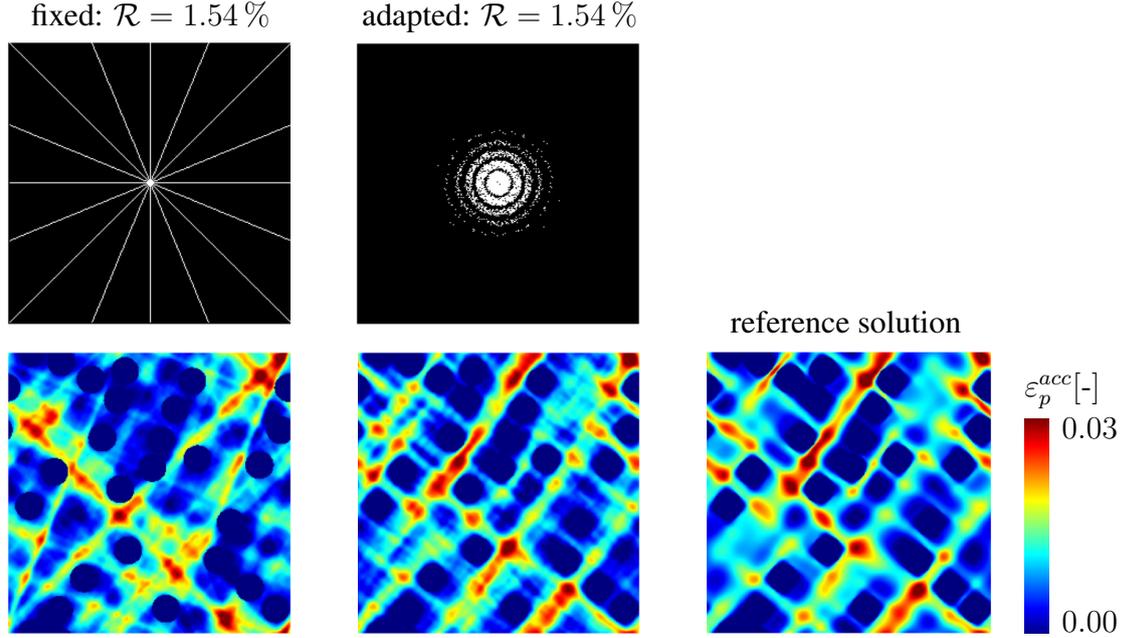

\begin{figure}[H]
	\hspace{-1.5cm}
	\centering
	\includegraphics[width=0.95\textwidth]{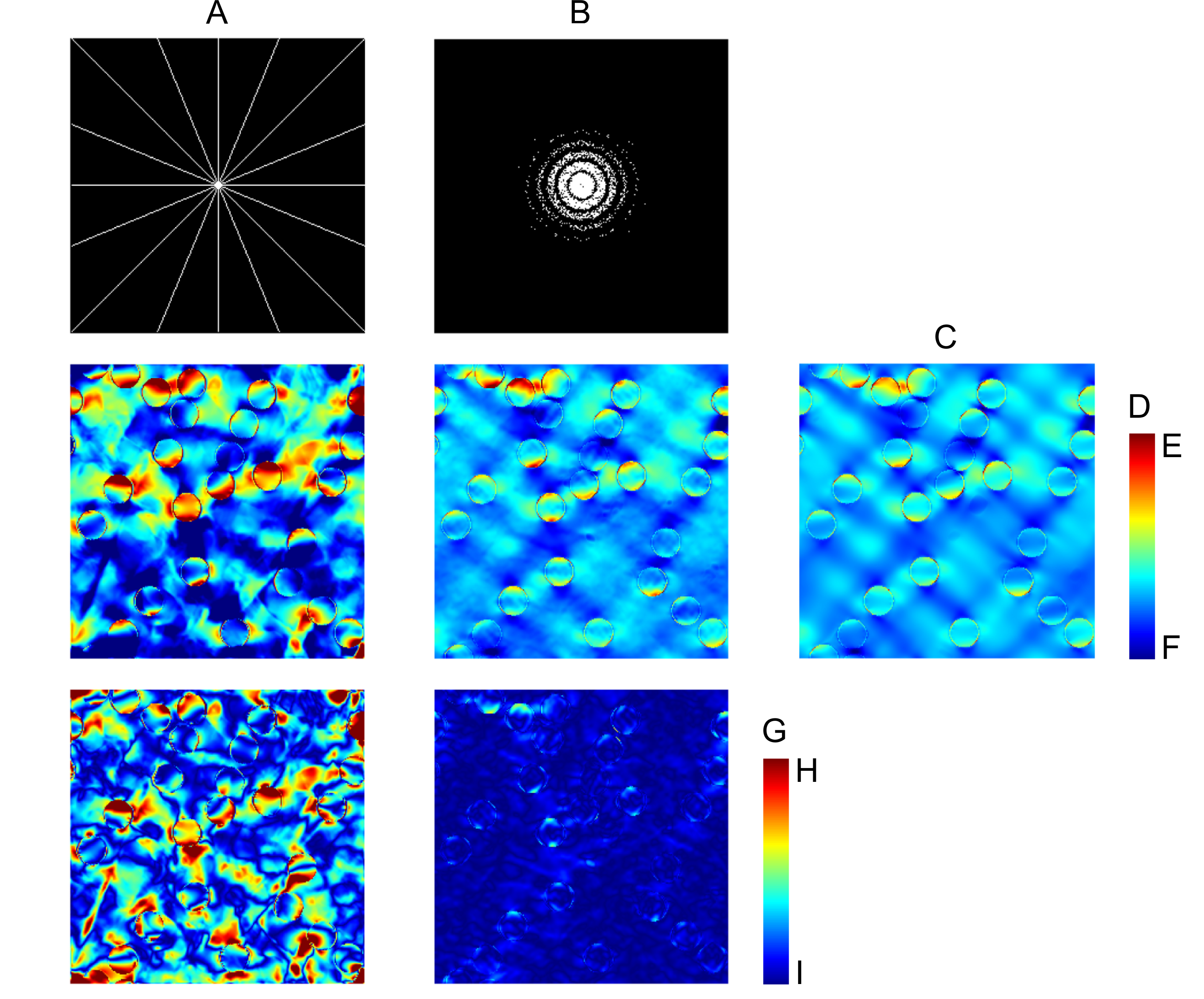}
	\put(-403,351){\colorbox{white}{fixed: $\mathcal R = 1.54\,\%$}}
	\put(-278,351){\colorbox{white}{adapted: $\mathcal R = 1.54\,\%$}}
	\put(-140,235){\colorbox{white}{reference solution}}
	\put(-30,210){\colorbox{white}{$\sigma_{11}[\text{GPa}]$}}
	\put(-16,195){\colorbox{white}{$0.02$}}
	\put(-16,122){\colorbox{white}{$0.00$}}
	\put(-162,94){\colorbox{white}{$\Delta\sigma_{11}[\text{GPa}]$}}
	\put(-148,78){\colorbox{white}{$0.01$}}
	\put(-148,5){\colorbox{white}{$0.00$}}
	\caption{Microstructural fields for the 2D elasto-plastic microstructure with several circular inclusions. \textit{Top row}: Fixed and adapted sampling pattern with the same number of wave vectors. \textit{Middle row}: Corresponding microstructural stress field $\sigma_{11}$ incorporating the reconstruction and compatibility step for the solution of the fixed sampling pattern and only the compatibility step for the solution of the adapted sampling pattern and reference stress field computed with the full set of frequencies. \textit{Bottom row}: Absolute difference in the microstructural stress field $\Delta\sigma_{11}$.}
	\label{fig:SeveralPlastComp}
\end{figure}

\begin{figure}[H]l
	\hspace{-1.5cm}
	\centering
	\includegraphics[width=0.95\textwidth]{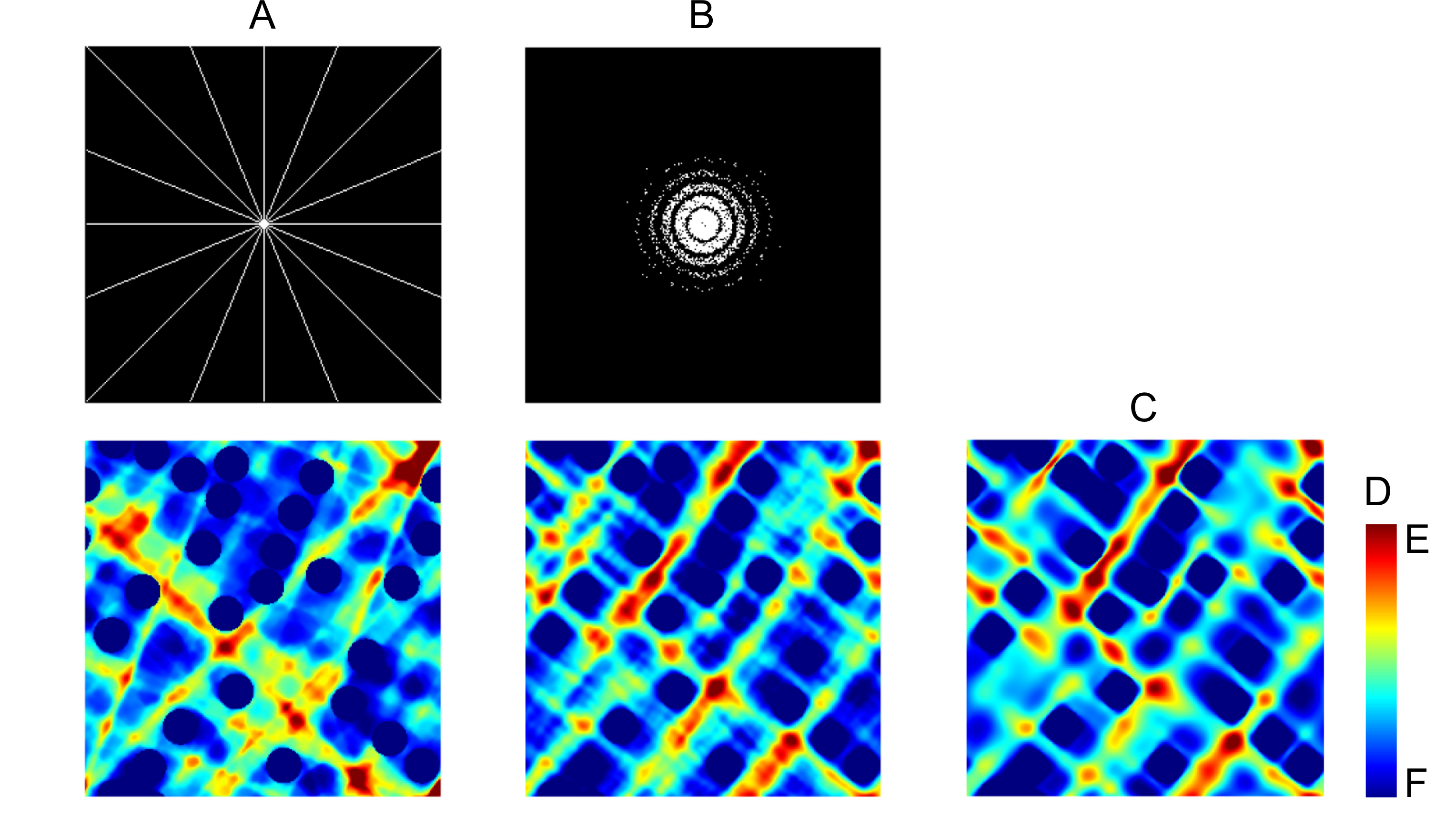}
	\put(-402,235){\colorbox{white}{fixed: $\mathcal R = 1.54\,\%$}}
	\put(-278,235){\colorbox{white}{adapted: $\mathcal R = 1.54\,\%$}}
	\put(-140,119){\colorbox{white}{reference solution}}
	\put(-30,95){\colorbox{white}{$\varepsilon_p^{acc}[\text{-}]$}}
	\put(-16,79){\colorbox{white}{$0.03$}}
	\put(-16,6){\colorbox{white}{$0.00$}}
	\caption{Microstructural fields for the 2D elasto-plastic microstructure with several circular inclusions. \textit{Top row}: Fixed and adapted sampling pattern with the same number of wave vectors. \textit{Bottom row}: Corresponding microstructural accumulated plastic strain field $\varepsilon_p^{acc}$ incorporating the reconstruction and compatibility step for the solution of the fixed sampling pattern and only the compatibility step for the solution of the adapted sampling pattern and reference accumulated plastic strain field computed with the full set of frequencies.}
	\label{fig:SeveralPlastAlphaComp}
\end{figure}
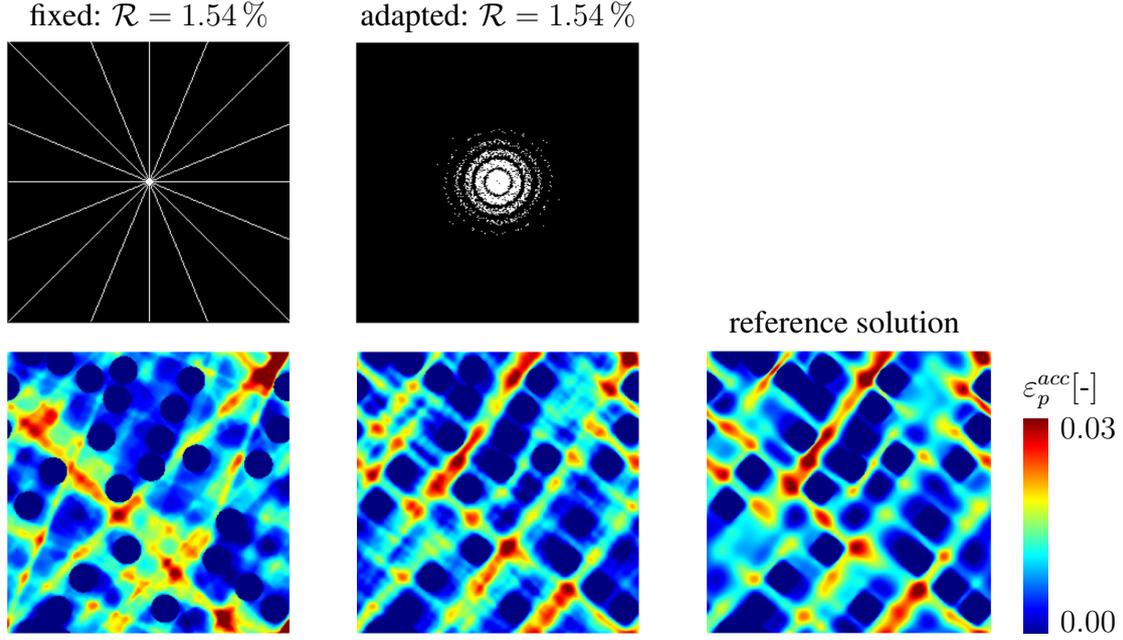

In the following, a composite with several circular inclusions is investigated. The elastic constants are the same as in the example before: $\lambda_\mathrm{I}=2.0 \, \text{GPa}$ and $\mu_\mathrm{I}=2.0 \, \text{GPa}$ for the inclusion and $\lambda_\mathrm{M}=1.0 \, \text{GPa}$ and $\mu_\mathrm{M}=1.0 \, \text{GPa}$ for the matrix material, respectively. Besides the elastic solution, also a simulation with an elasto-plastic matrix material behavior is performed. Considering an elasto-plastic material behavior of the matrix, the additional material parameters are set to $H_\mathrm{M}=0.01 \, \text{GPa}$ as hardening modulus and $\sigma_\mathrm{yM}^0=0.01\,\text{GPa}$ as initial yield stress. The investigated microstructure and the prescribed macroscopic strain is presented in Figure \ref{fig:MicroSeveral}.
For the fixed and the adapted sampling pattern the same amount of frequencies is used. Figure \ref{fig:SeveralRedu} shows the microstructural stress fields $\sigma_{11}$ as well as the differences $\Delta\sigma_{11}$ to the reference solution for the pure elastic case. As already seen for one inclusion, the error in the solution based on the adapted reduced set of frequencies is significantly lower compared to the solution of the fixed sampling pattern. Regarding the solution with the adapted set of frequencies, the highest differences occur again in the transition from matrix to inclusion. Instead, for the fixed sampling pattern, the highest errors occur within the inclusions.
Again, we perform the reconstruction and the compatibility step for the solution based on the fixed sampling pattern and only the compatibility step for the adapted sampling pattern. The corresponding microstructural fields are shown in Figure \ref{fig:SeveralComp}. Here, similar effects as described in Chapter \ref{chap:OneInc} for a microstructure with only one inclusion occur: The solution with the adapted sampling pattern is more accurate.
Since the solution behavior of the elastic microstructures with one or several inclusions is similar, we do not present further results on that and focus in the following on the microstructure with elasto-plastic material behavior. \\
The generation of a geometrically adapted sampling pattern does not depend on the material behavior itself, but only on the geometrical representation of the matrix and inclusions. Due to that, the geometrically adapted sampling pattern for the microstructure with several elastic inclusions and an elasto-plastic matrix material behavior is the same as for the microstructure with several inclusions and an overall elastic material behvior. Figure \ref{fig:SeveralPlastRedu} shows the microscopic stress field $\sigma_{11}$ corresponding to the fixed and adapted sampling pattern, the reference solution and the absolute difference in the reduced solution compared to the reference solution $\Delta \sigma_{11}$. It can be seen, that the stress difference for the fixed and the adapted sampling pattern is in general higher considering the nonlinear matrix material behavior instead of the purely linear material behavior shown in Figure \ref{fig:SeveralRedu}. Nevertheless, the error in the solution with the adapted sampling pattern is again significantly lower than the error in the solution with the fixed sampling pattern. In addition, the error considering the adapted sampling pattern is still related to the transition from inclusion to matrix material, while the error for the fixed sampling pattern is again particularly high within the inclusions.  
Figure \ref{fig:ErrorPlast} shows the macroscopic error $\bar{\mathcal E}$ (\textit{left}) and the microscopic error $\mathcal E$ (\textit{right}) again based on the reduced set of frequencies $\mathcal R$ for the solution with the fixed and adapted sampling pattern for the elasto-plastic composite. Incorporating $\mathcal R= 1.54\,\%$ and considering the fixed sampling pattern, these errors read $\bar{\mathcal E}\approx 34\,\%$ and $\mathcal E\approx 79\,\%$. Using the same amount of frequencies and the adapted sampling pattern, the errors reduce to $\bar{\mathcal E}\approx 2\,\%$ and $\mathcal E\approx 14\,\%$, respectively. In addition, Figure \ref{fig:ErrorPlast} shows that at some point ($\mathcal R \approx 15 \,\%$) the fixed sampling pattern leads to better results than the adapted sampling pattern. This might be related to the elasto-plastic material behavior of the matrix which results in a material behavior which is not that uniform within the matrix as the pure elastic material behavior. The adapted sampling pattern is only related to the geometrical representation of the matrix material and does not represent the different elasto-plastic material states. Due to that, the fixed sampling pattern, which is not bounded to the geometry, may lead to better results for a high amount of frequencies. Besides that, in the range of interest with a highly reduced set of frequencies, the adapted sampling pattern is performing significantly better than the fixed sampling pattern.
In Figure \ref{fig:SeveralPlastAlphaRedu} the accumulated plastic strain field $\varepsilon_p^{acc}$ for $\mathcal R=1.54\,\%$ of frequencies is shown. Also here, significant differences of the solution with the fixed sampling pattern compared to the reference solution are observed. For example, the accumulated plastic strain in the upper middle has a value of $\varepsilon_p^{acc}\approx0.3$ in the reference solution, while the accumulated plastic strain in the reduced solution has a value of $\varepsilon_p^{acc}\approx0$ at the same position. In contrast to that, the solution based on the adapted sampling pattern and the same amount of frequencies is quite similar to the reference solution.
Incorporating the reconstruction and compatibility step for the solution of the fixed sampling pattern and only the compatibility step for the solution of the adapted sampling pattern leads to the results given in Figures \ref{fig:SeveralPlastComp} and \ref{fig:SeveralPlastAlphaComp}. Figure \ref{fig:SeveralPlastComp} shows the microstructural stress field $\sigma_{11}$  and Figure \ref{fig:SeveralPlastAlphaComp} shows the accumulated plastic strain field $\varepsilon_p^{acc}$, respectively. Considering the fixed sampling pattern, it can be seen, that the calculated stress within the inclusions is improved by the reconstruction and the compatibility step, while the stress within the elasto-plastic matrix is not improved significantly. This is related to the accumulated plastic strain field, shown in Figure \ref{fig:SeveralPlastAlphaComp}, which is also not improved by these post-processing steps. As shown in Figure \ref{fig:SeveralPlastComp}, the microstructural stress field related to the solution with the adapted sampling pattern is slightly improved by solving the Lippmann-Schwinger equation once with the full set of frequencies.
\vspace{-0.1cm}
The CPU times for both sampling patterns and the different amount of frequencies are given in Tables \ref{tab:PlastTimeFixed} and \ref{tab:PlastTimeAdapted}. The behavior of the CPU times is similar compared to the elastic case shown in Tables \ref{tab:OneIncTimeFixed} and \ref{tab:OneIncTimeAdapted}, while a speed-up factor of 5 - 6 is gained by considering a set of $\mathcal R = 1.54 \,\%$ of frequencies in the nonlinear case.
\begin{table}[H]
	\centering
	\begin{tabular}{|c||c|c|c||c||c|}
		\hline
		elasto-plastic & \multicolumn{5}{c|}{\textbf{CPU time [s] - fixed sampling pattern}}\\ \hline \hline
		\(\mathcal{R}\,[\%]\) & total & \(-\hat{\BbbGammaVar}^{(0)}\hat{\bm{\tau}}(\bm{\varepsilon})\) (mean) & \(\bm{\sigma}(\bm{\varepsilon})\) (mean) & reconstruction & compatibility \\
		&&$t=100$ & $t=100$ & &\\
		\hline
		1.54  & 316.7  & 0.009 & 0.022 & 91.134  & 0.334 \\ \hline
		3.06  & 446.9  & 0.015 & 0.028 & 105.043 & 0.442 \\ \hline
		6.02  & 555.9  & 0.027 & 0.025 & 76.939  & 0.327 \\ \hline
		11.64 & 797.9  & 0.050 & 0.027 & 73.645  & 0.340 \\ \hline
		21.66 & 1139.4 & 0.087 & 0.024 & 65.870  & 0.339 \\ \hline
		36.79 & 1645.2 & 0.146 & 0.024 & 60.556  & 0.321 \\ \hline 
		\vdots & \vdots & \vdots & \vdots & \vdots & \vdots \\ \hline
		unreduced & 1866.5 & 0.241 & 0.024 & - & -\\ \hline
	\end{tabular}
	\caption{Total CPU time with mean CPU time per iteration step for solving the convolution integral and the constitutive law and CPU times for the reconstruction and the compatibility step of the simulation with the fixed sampling pattern for the 2D elasto-plastic microstructure with several circular inclusions.}
	\label{tab:PlastTimeFixed}
\end{table}
\begin{table}[H]
	\centering
	\begin{tabular}{|c||c|c|c||c||c|}
		\hline
		elasto-plastic & \multicolumn{5}{c|}{\textbf{CPU time [s] - adapted sampling pattern}}\\ \hline \hline
		\(\mathcal{R}\,[\%]\) & total & \(-\hat{\BbbGammaVar}^{(0)}\hat{\bm{\tau}}(\bm{\varepsilon})\) (mean) & \(\bm{\sigma}(\bm{\varepsilon})\) (mean) & reconstruction & compatibility\\ 
		&&$t=100$ & $t=100$ & &\\
		\hline
		1.54  & 366.7  & 0.008  & 0.025& - & 0.323 \\ \hline
		3.06  & 431.0  & 0.015  & 0.020& - & 0.342 \\ \hline
		6.02  & 533.8  & 0.026  & 0.021& - & 0.334 \\ \hline
		11.64 & 744.4  & 0.050  & 0.023& - & 0.338 \\ \hline
		21.66 & 1081.5 & 0.087  & 0.020& - & 0.324 \\ \hline
		36.79 & 1637.3 & 0.149  & 0.023& - & 0.342 \\ \hline 
		\vdots & \vdots & \vdots & \vdots & \vdots & \vdots \\ \hline
		unreduced & 1866.5 & 0.241 & 0.024 & - & -\\ \hline
	\end{tabular}
	\caption{Total CPU time with mean CPU time per iteration step for solving the convolution integral and the constitutive law and CPU times for the reconstruction and the compatibility step of the simulation with the adapted sampling pattern for the 2D elasto-plastic microstructure with several circular inclusions.}
	\label{tab:PlastTimeAdapted}
\end{table}

\subsection{Elastic 3D two phase material with one inclusion}
\label{chap:3D}
\begin{figure}[H]
	\hspace{-1.5cm}
	\centering
	\includegraphics[width=0.75\textwidth]{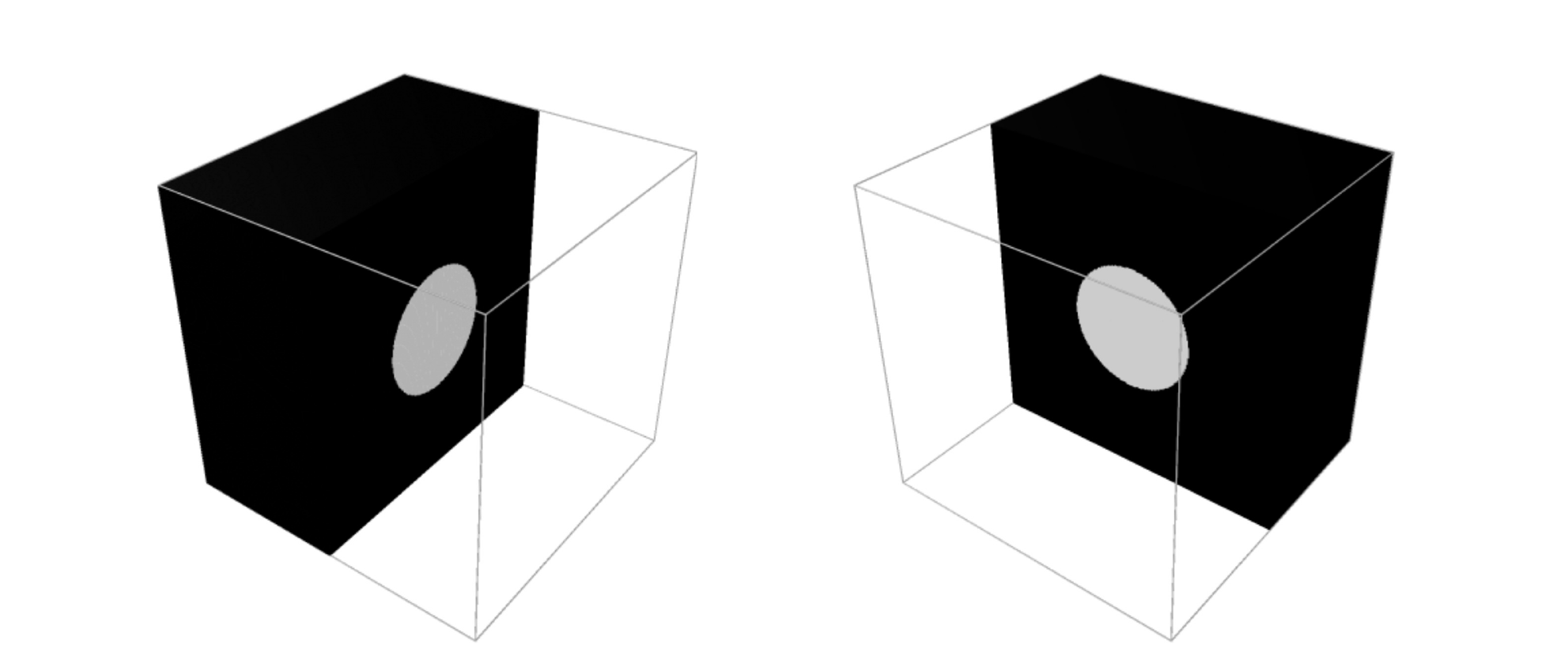}
	\caption{Two different intersections of the 3D microstructure with one spherical inclusion.}
	\label{fig:3DMicro}
\end{figure}

\begin{figure}[H]
	\hspace{-1.5cm}
	\centering
	\includegraphics[width=0.95\textwidth]{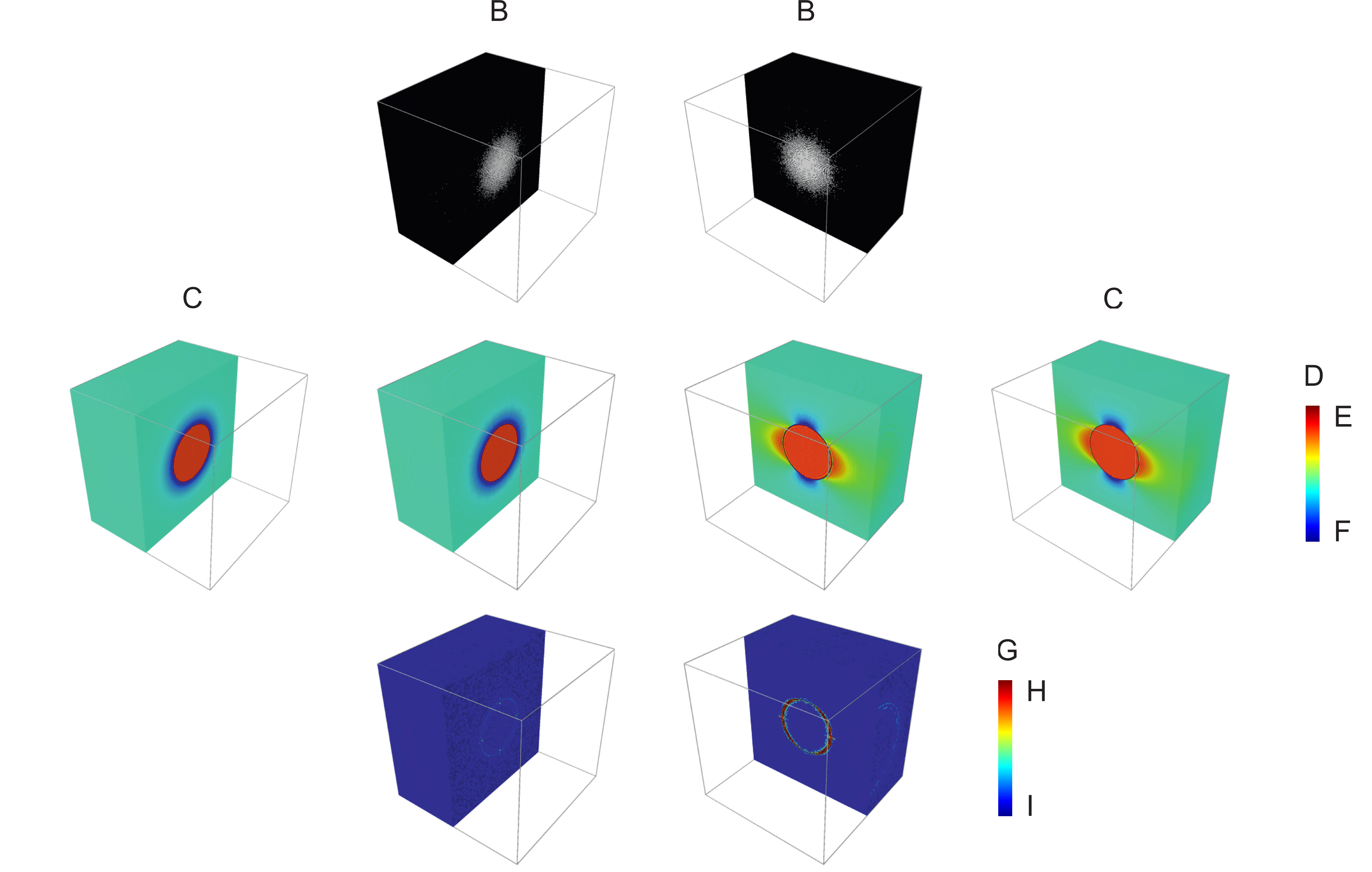}
	\put(-313,270){\colorbox{white}{$\mathcal R = 1.54\,\%$}}
	\put(-213,270){\colorbox{white}{$\mathcal R = 1.54\,\%$}}
	\put(-420,180){\colorbox{white}{reference solution}}
	\put(-130,180){\colorbox{white}{reference solution}}
	\put(-25,157){\colorbox{white}{$\sigma_{11}[\text{GPa}]$}}
	\put(-16,140){\colorbox{white}{$0.045$}}
	\put(-16,104){\colorbox{white}{$0.020$}}
	\put(-122,69){\colorbox{white}{$\Delta\sigma_{11}[\text{GPa}]$}}
	\put(-113,53){\colorbox{white}{$0.002$}}
	\put(-113,18){\colorbox{white}{$0.000$}}
	\caption{Microstructural fields of two different intersections for the 3D elastic microstructure with one circular inclusion. \textit{Top row}: Adapted sampling pattern. \textit{Middle row}: Corresponding microstructural stress field $\sigma_{11}$ and reference stress field computed with the full set of frequencies. \textit{Bottom row}: Absolute difference in the microstructural stress field $\Delta\sigma_{11}$.}
	\label{fig:3DSig}
\end{figure}
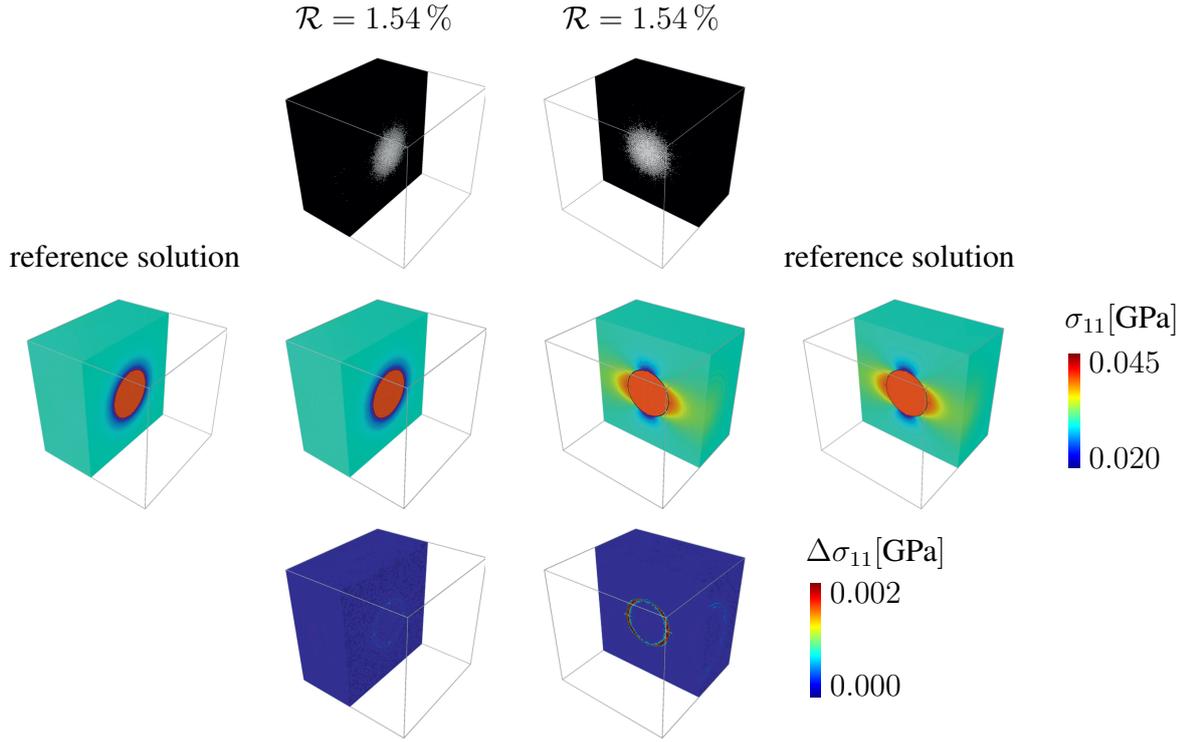

\begin{figure}[H]
	\includegraphics[width=\textwidth]{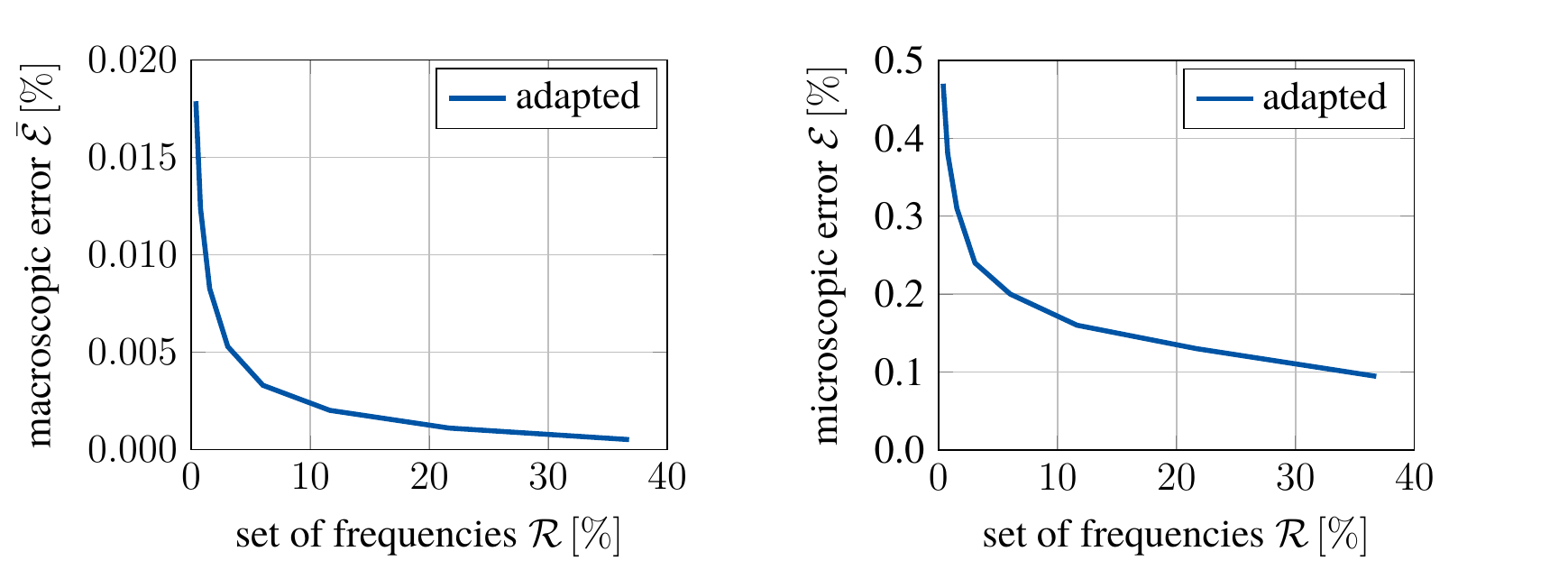}
	\caption{Macroscopic error $\bar{\mathcal E}$ (\textit{left}) and microscopic error $\mathcal E$ (\textit{right}) for the 3D elastic microstructure with one circular inclusion depending on the percentage of used frequencies $\mathcal R$ for the solution with the adapted sampling pattern.}
	\label{fig:Error3D}
\end{figure}

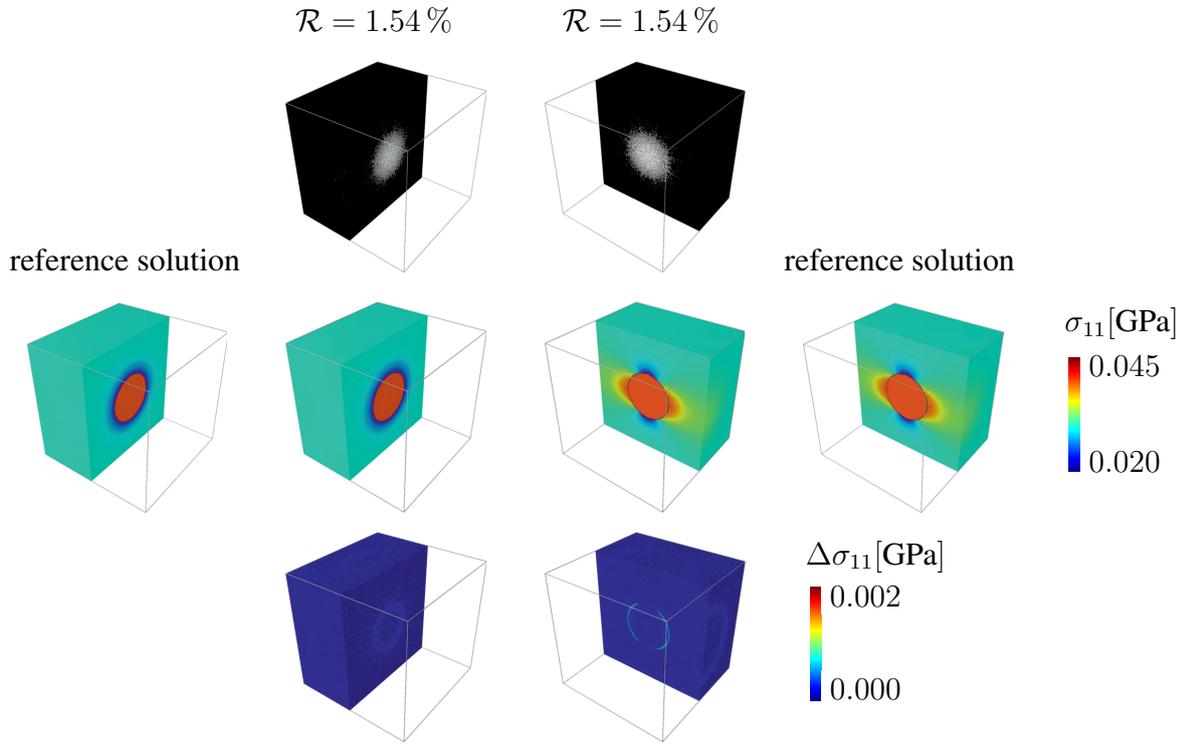
\begin{figure}[H]
	\hspace{-1.5cm}
	\centering
	\includegraphics[width=0.95\textwidth]{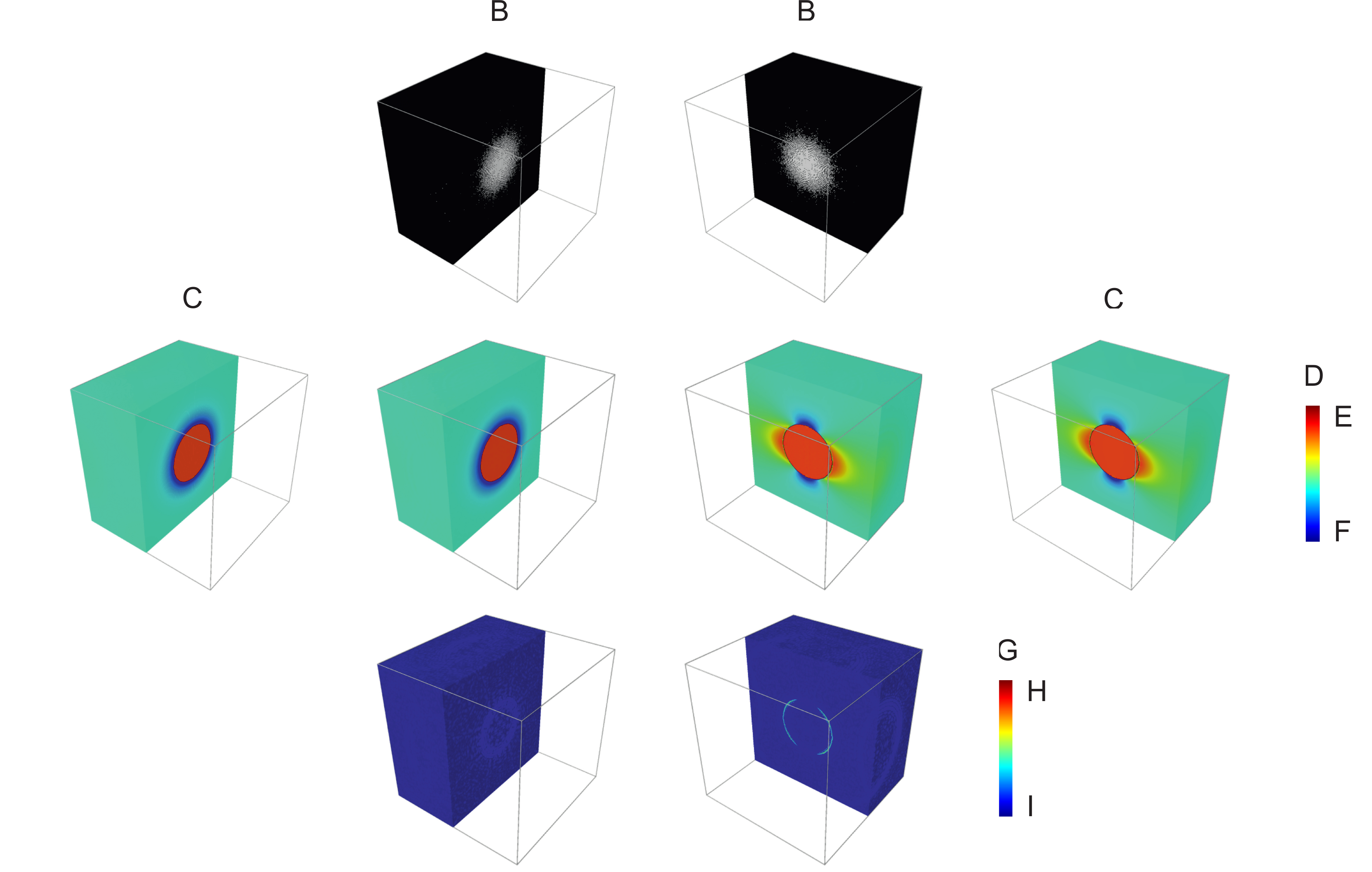}
	\put(-313,270){\colorbox{white}{$\mathcal R = 1.54\,\%$}}
	\put(-213,270){\colorbox{white}{$\mathcal R = 1.54\,\%$}}
	\put(-420,180){\colorbox{white}{reference solution}}
	\put(-130,180){\colorbox{white}{reference solution}}
	\put(-25,157){\colorbox{white}{$\sigma_{11}[\text{GPa}]$}}
	\put(-16,140){\colorbox{white}{$0.045$}}
	\put(-16,104){\colorbox{white}{$0.020$}}
	\put(-122,69){\colorbox{white}{$\Delta\sigma_{11}[\text{GPa}]$}}
	\put(-113,53){\colorbox{white}{$0.002$}}
	\put(-113,18){\colorbox{white}{$0.000$}}
	\caption{Microstructural fields of two different intersections for the 3D elastic microstructure with one circular inclusion. \textit{Top row}: Adapted sampling pattern. \textit{Middle row}: Corresponding microstructural stress field $\sigma_{11}$ incorporating the compatibility step and reference stress field computed with the full set of frequencies. \textit{Bottom row}: Absolute difference in the microstructural stress field $\Delta\sigma_{11}$.}
	\label{fig:3DRecon}
\end{figure}

Finally, a 3D microstructure with one centered spherical inclusion, as shown in Figure \ref{fig:3DMicro}, is investigated. The inclusions and the matrix material are considerd to be elastic again with $\lambda_\mathrm{I}=2.0 \, \text{GPa}$ and $\mu_\mathrm{I}=2.0 \, \text{GPa}$ for the inclusion and $\lambda_\mathrm{M}=1.0 \, \text{GPa}$ and $\mu_\mathrm{M}=1.0 \, \text{GPa}$ considering the matrix material, respectively. The applied macroscopic strain is set to $\bar\varepsilon_{11}=0.01$.
To generate the geometrically adapted sampling pattern in the 3D case, the same strategy as in the 2D case is used: First, the geometry is represented by a 3D step function. This function is transferred into Fourier space. The set of frequencies with the highest amplitudes for the representation of the step function in Fourier space is used for the reduced simulation. The resulting adapted sampling pattern for $\mathcal R =1.54\,\%$ of frequencies and the corresponding microstructural results are given in Figure \ref{fig:3DSig}. Since the 3D case with a spherical inclusion corresponds to the 2D case with a circular inclusion, the resulting sampling patterns are similar, see Figure \ref{fig:OneCircAdapted}. Also the behavior of the microstructural fields is similar to the 2D case described in Chapter \ref{chap:OneInc}. Using only $\mathcal R =1.54\,\%$ of the frequencies, the micromechanical stress fields of the reduced simulation already match the reference solution very well. The highest errors are again within the transition from matrix to inclusion.   
The macroscopic error $\bar{\mathcal E}$ and the microscopic error $\mathcal E$ depending on the reduced set of frequencies $\mathcal R$ for the 3D case are plotted in Figure \ref{fig:Error3D}. 
Using the compatibility step, the microstructural stress field $\sigma_{11}$ is again improved as shown in Figure \ref{fig:3DRecon}.
Table \ref{tab:3DTimeAdapted} shows the CPU times for the reduced and reference solution in the 3D case. It can be seen, that a significant speed up factor of up to approximately 100 is gained by using the geometrically adapted reduced set of frequencies with $\mathcal R = 1.54\,\%$ of frequencies. This speed up is again only gained by solving the convolution integral in Fourier space with the reduced set of frequencies.
\begin{table}[H]
	\centering
	\begin{tabular}{|c||c|c|c||c||c|}
		\hline
		3D composite & \multicolumn{5}{c|}{\textbf{CPU time [s] - adapted sampling pattern}}\\ \hline \hline
		\(\mathcal{R}\,[\%]\) & total & \(-\hat{\BbbGammaVar}^{(0)}\hat{\bm{\tau}}(\bm{\varepsilon})\) (mean) & \(\bm{\sigma}(\bm{\varepsilon})\) (mean) & reconstruction & compatibility\\ 
		\hline
		0.39  &  301.46 &    2.21 & 13.998  & - & 279.1\\ \hline
		0.78  &  284.79 &    3.32 & 13.960  & - & 284.9\\ \hline
		1.54  &  305.54 &    5.36 & 13.981  & - & 298.7\\ \hline
		3.06  &  351.76 &    9.59 & 14.411  & - & 277.2\\ \hline
		6.02  &  426.57 &   17.66 & 13.826  & - & 278.7\\ \hline
		11.64 &  584.64 &   32.94 & 14.368  & - & 278.1\\ \hline
		21.66 &  853.26 &   60.02 & 14.149  & - & 281.3 \\ \hline
		36.79 & 1260.20 &  100.94 & 13.920  & - & 279.8 \\ \hline
		\vdots & \vdots & \vdots & \vdots & \vdots & \vdots \\ \hline
		unreduced & 28188.0 & 2795.6 & 12.241 & - & -\\ \hline
	\end{tabular}
	\caption{Total CPU time with mean CPU time per iteration step for solving the convolution integral and the constitutive law and CPU times for the reconstruction and the compatibility step of the simulation with the adapted sampling pattern for the 3D elastic microstructure with one circular inclusion.}
	\label{tab:3DTimeAdapted}
\end{table}

\section{Conclusion and outlook}
\label{sec:summary}
We presented a novel approach to identify a sampling pattern for a reduced set of frequencies which is used for the FFT-based microstructure simulation. The approach is based on transferring the microstructural phase distribution represented by a step function into Fourier space and identifying the corresponding frequencies with the highest amplitudes. A given percentage of these frequencies with the highest amplitudes is subsequently used to determine the reduced set of frequencies. As shown for several two and three dimensional examples, such an adapted sampling pattern leads to significant better microstructural and overall results compared to the earlier introduced fixed sampling pattern. Considering only a few frequencies in the reduced wave vector, the error compared to the reference solution is so small, that a reconstruction is not necessary anymore. Therefore, the solution algorithm is in addition much easier, especially in the 3D case. Using the proposed solution strategy, a speed up factor of 5-8 in the 2D case and up to approximately 100 in the 3D case is obtained. \\
The proposed solution strategy only reduces the computational effort of solving the convolution in Fourier space, so that the next step is the combination of the proposed MOR technique with e.g. a clustering analysis, since the most time consuming part for the simulation of a complex material behavior is the evaluation of the material law. Using such a microstructural clustering analysis the stress evaluation in each grid point is reduced to a stress evaluation in a defined number of clusters instead. Based on previous works \citep{Wulfinghoff2017,Cavaliere2020,Waimann21} an additional significant speed-up is expected by combining both methods.\\

%%%%%%%%%%%%%%%%%%%%%%%%%%%%%%%%%%%%%%%%%%%%%%%%%%%%%%%%%%%%%%%%%%%%%%%%%%%%%%%%%%%%%%%%%%%%%%%%%%%%%%%%%

\textit{Acknowledgements}:~The authors gratefully acknowledge the financial support of the research work by the German Research Foundation (DFG, Deutsche Forschungsgemeinschaft) within the transregional Collaborative Research Center SFB/TRR 136, project  number 223500200, subproject M03. In addition Stefanie Reese gratefully acknowledges the financial support of the research work by the German Research Foundation (DFG, Deutsche Forschungsgemeinschaft) within the transregional Collaborative Research Center SFB/TRR 280, project number 417002380, subproject A01 and the  project ``Model order reduction in space and parameter dimension - towards damage-based modeling of polymorphic uncertainty in the context of robustness and reliability'', project  number 312911604, from the  priority  program (SPP) 1886.

\bibliographystyle{agsm}
\bibliography{literature}

\end{document}